\documentclass[11pt,a4paper]{article}
\usepackage{geometry}
\usepackage{multirow}
\usepackage[T1]{fontenc}
\usepackage[utf8]{inputenc}
\usepackage[english]{babel}
\usepackage{amsmath}
\usepackage{amsthm}
\usepackage{amsfonts}
\usepackage{mathtools}
\usepackage{amssymb} 
\usepackage{indentfirst}
\usepackage{graphicx}

\usepackage{afterpage}
\usepackage{mathrsfs}
\usepackage{subfig}
\usepackage{float}
\usepackage{xcolor}
\usepackage{url}
\usepackage{todonotes}
\usepackage{url, enumitem} 			
\usepackage[noadjust]{cite}				
\usepackage{epigraph}						
\usepackage[nottoc]{tocbibind}	

\title{Population dynamics of multiple ecDNA types}
\author{Elisa Scanu$^{2}$, Benjamin Werner$^{3}$, Weini Huang$^{1,2,*}$  \\
        \small $^{1}$ Group of Theoretical Biology, School of Life
        Science, Sun Yat-sen University,\\
        \small Guangzhou, China \\
        \small $^{2}$ School of Mathematical Sciences, Queen Mary University of London, \\
        \small United Kingdom \\
        \small $^{3}$ Centre for Cancer Genomics and Computational Biology, Barts Cancer Centre, \\
        \small Queen Mary University of London, United Kingdom\\
        \small \\
        \small $^*$ Corresponding author: weini.huang@qmul.ac.uk}
\date{April 2025}

\begin{document}
\maketitle

\begin{abstract}
Extrachromosomal DNA (ecDNA) can drive oncogene amplification, gene expression and intratumor heterogeneity, representing a major force in cancer initiation and progression. The phenomenon becomes even more intricate as distinct types of ecDNA present within a single cancer cell. While exciting as a new and significant observation across various cancer types, there is a lack of a general framework capturing the dynamics of multiple ecDNA types theoretically. Here, we present novel mathematical models investigating the proliferation and expansion of multiple ecDNA types in a growing cell population. By switching on and off a single parameter, we model different scenarios including ecDNA species with different oncogenes, genotypes with same oncogenes but different point mutations and phenotypes with identical genetic compositions but different functions. We analyse the fraction of ecDNA-positive and free cells as well as how the mean and variance of the copy number of cells carrying one or more ecDNA types change over time. Our results showed that switching does not play a role in the fraction and copy number distribution of total ecDNA-positive cells, if selection is identical among different ecDNA types. In addition, while cells with multiple ecDNA cannot be maintained in the scenario of ecDNA species without extra fitness advantages, they can persist and even dominate the ecDNA-positive population if switching is possible.

\end{abstract}

\section*{Introduction}

Cancer is the uncontrolled growth of abnormal cells, and it develops from a single cell to bigger masses which take control of portions of healthy tissues \cite{1,2}. Following the Darwinian evolution theory, in order to successfully invade, tumour cells obtaining simple or complex variations can survive and escape the normal control mechanism of tissues \cite{3, 37, 38}. One efficient way is through copy number alteration, including amplification of functional genomics sequences such as oncogenes \cite{1,4}. The amplification of specific oncogenes can happen through extra chromosomal DNA (ecDNA) and homogeneously staining regions (HSRs), which are thought to be the cytogenetic hallmarks of gene amplification in cancers \cite{6}. There is much research related to HSRs \cite{6,5}, but the role of ecDNA in driving tumor expansion and treatment resistance has been largely underestimated for many years. Recently, many studies suggested that patients with tumours containing ecDNA have worse clinical outcomes \cite{8,19}.  "Hiding" oncogenes within ecDNA fragments is a strategic mechanism for the evolution of cancer cells, as this relocation effectively liberates these elements from the normal chromosomal constraints like centromeres, consequently giving tumors the ability to undergo accelerated evolution compared to chromosomal genomes.

The observed ecDNAs are relatively large (1.3 Mb on average) and highly amplified circular structures containing genes and regulatory regions \cite{1}, and they make unique contributions to oncogenic transcription and tumour progression through their promotion of oncogene amplification and treatment resistance \cite{19,13,14,15,16,17}. The origin of ecDNA has been long studied, and several possible mechanisms have been proposed, including the breakage-fusion-bridge (BFB) cycle, translocation-deletion-amplification model, “episome” model and chromothripsis \cite{18,31}. The segregation patterns of ecDNA elements in cell divisions have been investigated throughfully, with the common conclusion that unlike DNA, ecDNA copies are randomly partitioned to the daughter cells from the mother \cite{19,21,22,23}, and this fuels a fast accumulation of intercellular copy number heterogeneity. A basic mathematical model of tumour dynamics with a single ecDNA type has already been developed \cite{22}, where the random segregation and driver role of ecDNA elements in tumour expansion has been validated through an integration of theory and experimental and clinical data \cite{22}. Here, we move beyond the single ecDNA type hypothesis and investigate the population dynamics of multiple ecDNA types.
 
There are multiple reasons and evidence why we need a general mathematical framework to model multiple ecDNA types, which can refer to  ecDNA species carrying distinct oncogenes, genotypes with the same oncogenes but different point mutations, and
phenotypes with identical genetic compositions but different functions generated by phenotypical switching. 
First, multiple ecDNAs species originally derived from different chromosomal loci have been observed in the same cancer cell \cite{24,29}. These ecDNAs can congregate in micron-sized hubs in the nucleus \cite{16,39} and enable intermolecular gene activation, where enhancer elements on one ecDNA molecule can activate coding sequences on another ecDNA \cite{29}. Second, ecDNAs own clustered somatic mutations \cite{27,28} which change the genotype of ecDNA and may alter the function of ecDNA sequences with potential phenotypic evolution like in resistance. Third, ecDNA is particularly sensitive to different epigenetic states, which have been associated with responses to site of growth \cite{25,26} and to stress, including hypoxia and high acidity encountered in tumour microenvironments \cite{22,24}, and activation or deactivation of different oncogenes. This means even without any genetic alternations, ecDNA might switch between different phenotypical states intrinsically. \\
Considering that experimental evidence shows the most prevalent presence of two ecDNA species in human cancer cells \cite{29}, we focus on our analysis of an evolutionary process with two ecDNA types. However, we can extend our general framework to model any number of ecDNA types. Specifically, we model the segregation, possible switching and selection of two ecDNA types in a growing population. Our investigation demonstrates that the stochastic segregation of extrachromosomal DNA yields diverse outcomes under selection and different switching scenarios. While selection significantly impacts the population composition of ecDNA-positive and ecDNA-free cells, as well as the mean, variance, and distribution of ecDNA copy numbers in the total population, switching only plays a role in these quantities when selection differs between the cells carrying different ecDNA types. However, if we classify the cell population into different subgroups based on composition of ecDNA types within a cell, we observe distinct dynamics of cells carrying different types of ecDNA by varying switching strengths. This holds even when selection is identical for different ecDNA types.
More specifically, we examined genotypes with extremely low switching rates or phenotypes with relatively higher switching rates, the mean copy number and proportion of cells in subgroups are significantly influenced by the switching rate. 

\section*{Methods}
\subsection*{A general framework of two ecDNA types}
\begin{figure}[H]
	\centering
	\includegraphics[width=\textwidth]{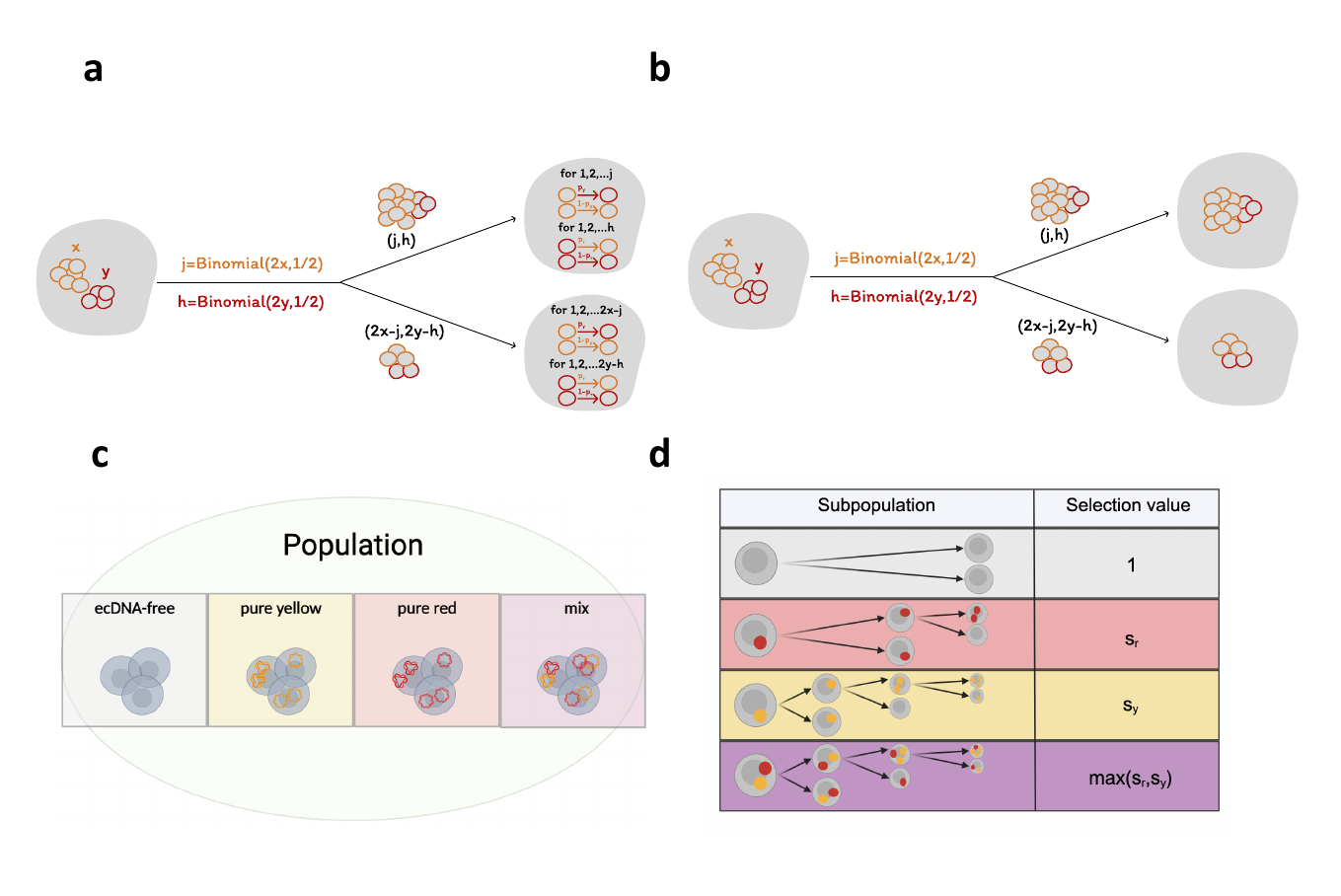}
	\caption{\textbf{A general framework for modeling two ecDNA types.} \textbf{a. Two ecDNA genotypes or phenotypes} through genetic mutations (small $p_y$, $p_r$) or phenotypical switching (large $p_y$, $p_r$). At each cell division, each type evolves undergoing independent binomial segregation and it can switch between the two ecDNA types with probabilities $p_y$ and $p_r$. \textbf{b. Two ecDNA species without switching ($p_y=p_r=0$).} When the p switching rates are zero, we just consider a random segregation of yellow and red ecDNA species in the daughter cells following two independent binomial distributions. \textbf{c. Cells subpopulations.} With or without switching, we can classify four subpopulations based on ecDNA types carried by cells, i.e. cells without ecDNA (ecDNA-free), cells with only one type of ecDNA (pure yellow and pure red) and cells carrying both ecDNA types (mix). \textbf{d Selection strengths.} We summarise the selection coefficients for different subpopulations. For the sake of simplification, we assume the mix cells have the same fitness as the maximum values of fitness of the pure yellow and pure red cells, which can be easily relaxed for an extension.}
	\label{modello}
\end{figure}

We consider two types of ecDNA, denoted as yellow and red respectively in Figure \ref{modello}. The reproduction rates (fitness in our model) of the two ecDNA types are $s_y$ and $s_r$, which are equal to $1$ if the selection between ecDNA-positive or ecDNA-free cells is neutral. This means that cells carrying ecDNA have the same fitness as cells without ecDNA under neutral selection, which is normalised to $1$. Alternatively, cells carrying ecDNA types can have fitness advantages compared to ecDNA-free cells, including two scenarios depending on whether the reproduction rates of carrying different ecDNA types equal ($s_y=s_r>1$) or not ($s_y\neq s_r>1$). For the sake of simplicity, we assume a cell carrying a mix of yellow and red ecDNA elements has a reproduction rate as the maximum value between $s_y$ and $s_r$ (Figure \ref{modello}d). This refers to an assumption that having both ecDNA types does not lead to an extra fitness advantage, which does not change our general framework and equations and can be easily released for an extension of analysis \cite{29}. 

During each cell division, the ecDNA copies are duplicated similar to the chromosomal genetic materials, but due to lack of centromeres the two types of ecDNA copies are randomly partitioned into the two daughter cells independently following a separate binomial distribution with the probability of a single copy segregated to any daughter cell as $1/2$. Note we can also release the assumption of the independent segregation between the two ecDNA species with evidence-based motivations \cite{29}. For our general framework, we focus on the analysis with independent segregation. 

We introduce two probabilities $p_y$ and $p_r$ in the range $[0,1]$, representing switching between ecDNA types coupled with cell divisions. If $p_y=p_r=0$, switching is off and the two ecDNA types refer to different ecDNA species (Figure \ref{modello}b),  where e.g. distinct oncogenes or enhancers are carried in the ecDNA elements and thus there is no one-step transition between the two ecDNA types. If at least one of $p_y$ and $p_r$ is greater than 0, then we consider the propagation of two ecDNA genotypes or phenotypes (Figure \ref{modello}a). Small values of $p_y$ and $p_r$ can refer to a genotypic change through e.g. point genetic mutations. With an ecDNA sequence size of $1$ Mb and a mutation rate of $10^{-9}$ to $10^{-8}$ per base pair, $p_y$ or $p_r$ could be in the order of $10^{-3}$ to $10^{-2}$. Larger values of $p_y$ and $p_r$ (but still far less than $1$) can be interpreted as intrinsic phenotype switching, where epigenetic changes such as through  DNA methylation \cite{19,36} or other molecular events alter the ecDNA phenotypes.

In a growing population, the random segregation and/or possible switching between the two ecDNA types promote a high heterogeneity among cells, which can be classified into four different subpopulations, i.e. pure yellow, pure red, mix and ecDNA-free cells (Figure \ref{modello}c$\&$d). We are interested in how the interplay between all evolutionary parameters $p_y, p_r, s_y, s_r$ will impact the dynamics of different subpopulations.

\subsection*{Deterministic dynamics}
These interactions can be expressed by two different mathematical approaches, which are a deterministic one comprising a system of ordinary differential equations (ODEs) and a demographic one comprising a system of stochastic equations. Starting from the system of ODEs, we analyse the densities of cells with and without ecDNA copies denoted by $n^+(t)$ and $n^-(t)$ respectively with,
\begin{equation}\label{sistemaweini}
    \begin{cases}
     \frac{d\,n^+(t)}{dt}=n^+(t)-\gamma (t)n^+(t)\,\\
    \frac{d\, n^-(t)}{dt}=n^-(t)+\gamma(t) n^+(t).
    \end{cases}
\end{equation}
The above system is justified by observing that cells carrying ecDNA can lose all ecDNA copies, but the reverse is not possible. Because the origin of ecDNA is often considered a random event due to  genomics instability, and generally a specific ecDNA can not be gained back once lost in a population. Moreover, the function $\gamma(t)$ corresponds to the at the moment undetermined loss rate of ecDNA due to complete asymmetric segregation, i.e. a daughter cell inherited all copies and the other one none by chance. 
In the case of two ecDNA types, we can split the density function $n^+(t)$ into the sum of three subfunctions: $\rho_{i,0}\big|_{i>0}(t)$, $\rho_{0,k}\big|_{k>0}(t)$ and $\rho_{i,k}\big|_{i,k>0}(t)$, respectively for pure yellow, pure red and mix cells:
\begin{equation}\label{sistemahomo}
\begin{cases}
\frac{d\, \rho_{i,0}(t)}{dt}=\rho_{i,0}(t)-\gamma(t) \rho_{i,0}(t)+ g_1(t) p_r \rho_{i,k}\big|_{i,k>0}(t) +k_1(t) p_r \rho_{0,k}(t) - h_2(t) p_y \rho_{i,0}(t) \\
\frac{d\, \rho_{0,k}(t)}{dt}=\rho_{0,k}(t)-\gamma(t) \rho_{0,k}(t)+ g_2(t) p_y \rho_{i,k}\big|_{i,k>0}(t) +h_1(t) p_y \rho_{i,0}(t) - k_2(t) p_y \rho_{0,k}(t) \\
\frac{d\, \rho_{i,k}(t)}{dt}\big|_{i,k>0}=\rho_{i,k}\big|_{i,k>0}(t)-\gamma(t) \rho_{i,k}\big|_{i,k>0}(t)- (g_1(t) p_r+g_2(t) p_y)\rho_{i,k}\big|_{i,k>0}(t) \\
+(h_2(t)-h_1(t)) p_y \rho_{i,0}(t) +(k_2(t)-k_1(t)) p_r \rho_{0,k}(t)
\end{cases}
\end{equation}
We can recover Eqs. (\ref{sistemaweini}) by summing the three equations in Eqs. (\ref{sistemahomo}) and generalise Eqs. (\ref{sistemahomo}) to positive selection cases by multiplying each subpopulation by its own selection coefficient, $s_y$, $s_r$ or $\max(s_y,s_r)$.\\

\subsection*{System of equations for the cells dynamics}

The ecDNA content of daughter cells at each division depends on the outcome of the random segregation and the  switching probability. We can break down all the possibilities of getting a cell with $i$ ecDNA copies upon division. The notation $C_{i,k}(t)$ indicates the number of cells with $i$ yellow and $k$ red copies at time $t$. Without loss of generality, we can set $s_y\ge s_r$, then the fitness advantage for mix cells is the maximum between $s_y$ and $s_r$, thus $s_y$ (Figure \ref{modello}).\\
We start by calculating the equation for pure yellow cells, i.e. $C_{i,0}(t)$, namely the number of cells with $i$ yellow copies at time t. This is:
\begin{equation}\label{yellow}
    \resizebox{.95\hsize}{!}{$
\begin{aligned}
    \frac{dC_{i,0}(t)}{dt}\bigg|_{i>0}&=\overbrace{-s_yC_{i,0}}^{\text{Mother divides}}\overbrace{+2s_y\sum_{j=\big[\frac{i}{2}  \big]}^{\infty} (1-p_y)^i C_{j,0} \binom{2j}{i}\frac1{2^{2j}}}^{\text{Pure yellow $\longrightarrow$ pure yellow}}\\
    &\overbrace{+2s_r\sum_{h=\big[\frac{i}{2}  \big]}^{\infty} p_r^i C_{0,h} \binom{2h}{i}\frac1{2^{2h}}}^{\text{Pure red $\longrightarrow$ pure yellow}}\\
    &\overbrace{+2s_y\sum_{\substack{j+h=\big[\frac{i}{2}\big]\\j, h>0 }}^{\infty} (1-p_y)^i C_{j,h} \binom{2j}{i}\binom{2h}{0}\frac1{2^{2(j+h)}}}^{\text{Mix $\longrightarrow$ pure yellow due to random segregation}}\\
    &\overbrace{+2s_y\sum_{\substack{j+h=\big[\frac{i}{2}\big]\\j, h>0 }}^{\infty}\sum_{v=0}^{i-1} \binom{i}{v} p_r^{i-v} (1-p_y)^{v}C_{j,h} \binom{2j}{v}\binom{2h}{i-v}\frac1{2^{2(j+h)}}}^{\text{Mix $\longrightarrow$ pure yellow}}\\
\end{aligned}
$}
\end{equation}
In the equation above, pure yellow cells can be obtained from pure yellow, pure red or mix mothers. If a pure yellow mother produces a daughter with $i$ yellow copies, then this enters the compartment $C_{i,0}$ if any copy switches colour ($(1-p_y)^i$). If a pure red mother produces a daughter with $i$ red copies, then this enters $C_{i,0}$ if all its copies switch ($p_r^i$). Finally, if a mix mother produces a daughter with $i$ copies, then: if all these are yellow, then the daughter becomes pure yellow if any of these copy switches; alternatively, if the daughter carries $v$ yellow and $i-v$ red copies, then it enters the compartment $C_{i,0}$ if all its red copies switch to yellow and all its yellow copies do not switch ($p_r^{i-v}(1-p_y)^v$. Notice how the last two terms of the equation can be merged in a unique sum by including the index $v=0$ in the last summation; however, we keep this terms separated in order to easily simplify this equation to the case of one-way switching, i.e. one between $p_y$ and $p_r$ is zero. This logic is summarised in Figure \ref{equation}a.\\
Similarly, the dynamics of pure red cells with $i$ ecDNA copies over time, i.e. $C_{0,i}(t)$, is:
\begin{equation}\label{red}
    \resizebox{.95\hsize}{!}{$
\begin{aligned}
    \frac{dC_{0,i}(t)}{dt}\bigg|_{i>0}&=\overbrace{-s_rC_{0,i}}^{\text{Mother divides}}\overbrace{+2s_y\sum_{j=\big[\frac{i}{2}  \big]}^{\infty} p_y^i C_{j,0} \binom{2j}{i}\frac1{2^{2j}}}^{\text{Pure yellow $\longrightarrow$ pure red}}\\
    &\overbrace{+2s_r\sum_{h=\big[\frac{i}{2}  \big]}^{\infty} (1-p_r)^i C_{0,h} \binom{2h}{i}\frac1{2^{2h}}}^{\text{Pure red $\longrightarrow$ pure red}}\\
    &\overbrace{+2s_y\sum_{\substack{j+h=\big[\frac{i}{2}\big]\\j, h>0 }}^{\infty} (1-p_r)^i C_{j,h} \binom{2j}{0}\binom{2h}{i}\frac1{2^{2(j+h)}}}^{\text{Mix $\longrightarrow$ pure red due to random segregation}}\\
    &\overbrace{+2s_y\sum_{\substack{j+h=\big[\frac{i}{2}\big]\\j, h>0 }}^{\infty}\sum_{v=1}^{i} \binom{i}{v} p_y^{v} (1-p_r)^{i-v}C_{j,h} \binom{2j}{v}\binom{2h}{i-v}\frac1{2^{2(j+h)}}}^{\text{Mix $\longrightarrow$ pure red}}\\
\end{aligned}
$}
\end{equation}
Pure red cells can be obtained from pure yellow, pure red or mix mothers similarly to equation (\ref{yellow}). The logic is summarised in Figure \ref{equation}b.\\
Finally, for the rate equation for mix cells with $k$ yellow and $i-k$ red copies at time $t$, i.e. $C_{k,i-k}(t)$, we have to distinguish three different cases, based on the value of the switching rates $p_y$ and $p_r$. If \textbf{both $p_y$ and $p_r$ are positive} (i.e. $p_y, p_r>0$), then the equation is:
\begin{equation}\label{mix}
    \resizebox{.95\hsize}{!}{$
\begin{aligned}
    \frac{dC_{k,i-k}(t)}{dt}\bigg|_{\substack{i,k>0 \\ k=1,\dots i-1}}&=\overbrace{-s_yC_{k,i-k}}^{\text{Mother divides}}\overbrace{+2s_y\sum_{j=\big[\frac{i}{2}\big]}^{\infty} \binom{i}{k} p_y^{i-k}(1-p_y)^{k}C_{j,0} \binom{2j}{i}\frac1{2^{2j}}}^{\text{Pure yellow $\longrightarrow$ mix ($k,i-k$)}}\\
    &\overbrace{+2s_r\sum_{h=\big[\frac{i}{2}\big]}^{\infty} \binom{i}{k} p_r^{k}(1-p_r)^{i-k}C_{0,h} \binom{2h}{i}\frac1{2^{2h}}}^{\text{Pure red $\longrightarrow$ mix ($k,i-k$)}}\\
    &\overbrace{+2s_y\sum_{\substack{j+h=\big[\frac{i}{2}  \big]\\j,h>0}}^{\infty}\sum_{v=1}^{i}\sum_{m=1}^{k} \binom{v}{m}\binom{i-v}{k-m} p_y^{v-m}(1-p_y)^{m}p_r^{k-m}(1-p_r)^{i-v-(k-m)}C_{j,h} \binom{2j}{v}\binom{2h}{i-v}\frac1{2^{2(j+h)}}}^{\text{Mix $\longrightarrow$ mix ($k,i-k$)}}\\
\end{aligned}
$}
\end{equation}
In the case where both $p_y, p_r>0$, mix cells with $k$ yellow and $i-k$ red ecDNA copies can be obtained from pure yellow, pure red or mix mothers. If a pure yellow mother produces a daughter with $i$ yellow copies, then this enters the compartment $C_{k,i-k}$ if $i-k$ copies switch to red and $k$ copies do not switch ($p_y^{i-k}(1-p_y)^k$). If a pure red mother produces a daughter with $i$ red copies, then this enters $C_{k,i-k}$ if $k$ copies switch to yellow and $i-k$ copies do not switch ($p_r^{k}(1-p_r)^{i-k}$). Finally, if a mix mother produces a daughter with $v$ yellow and $i-v$ red copies, then a cell in $C_{k,i-k}$ can be obtained by switching: if $v-m$ copies switch from yellow to red, where $m\in[1,\min(k,v)]$, then accordingly $k-m$ copies from the red ones should switch to yellow ($p_y^{v-m}(1-p_y)^{m}p_r^{k-m}(1-p_r)^{i-v-(k-m)}$). This logic is summarised in Figure \ref{equation}c.\\
Given the complexity of the master equation for this system, merging all the possible switching cases in a compact unique expression is challenging. Equations (\ref{yellow}) and (\ref{red}) can be adapted to the one-way switching case, i.e. either $p_y=0$ or $p_r=0$, just by substitution of the corresponding values for $p_y$ and $p_r$. Conversely, equation (\ref{mix}) needs a different version in order to correctly account for all the contributions from the different subpopulations in case of one-way switching. We then propose two version for equation (\ref{mix}) when one between $p_y$ or $p_r$ is zero.\\
If \textbf{$p_y$ is positive but $p_r$ is zero}, i.e. $p_y>0, p_r=0$, then the equation for $C_{k,i-k}(t)$ is:
\begin{equation}\label{mixoneway1}
    \resizebox{.95\hsize}{!}{$
\begin{aligned}
    \frac{dC_{k,i-k}(t)}{dt}\bigg|_{\substack{i,k>0 \\ k=1,\dots i-1}}&=-s_yC_{k,i-k}\overbrace{+2s_y\sum_{j=\big[\frac{i}{2}\big]}^{\infty}\binom{i}{k} p_y^{i-k}(1-p_y)^{k}C_{j,0} \binom{2j}{i}\frac1{2^{2j}}}^{\text{Pure yellow $\longrightarrow$ mix ($k,i-k$)}}\\
    &\overbrace{+2s_y\sum_{\substack{j+h=\big[\frac{i}{2}  \big]\\j=\big[\frac{k}{2}\big],h>0}}^{\infty}\sum_{v=k}^{i} \binom{v}{k} p_y^{v-k}(1-p_y)^{k}C_{j,h} \binom{2j}{v}\binom{2h}{i-v}\frac1{2^{2(j+h)}}}^{\text{Mix $\longrightarrow$ mix ($k,i-k$)}},\\
\end{aligned}
$}
\end{equation}
whilst in the case where \textbf{$p_r$ is positive but $p_y$ is zero}, i.e. $p_y=0, p_r>0$, we have:
\begin{equation}\label{mixoneway2}
    \resizebox{.95\hsize}{!}{$
\begin{aligned}
    \frac{dC_{k,i-k}(t)}{dt}\bigg|_{\substack{i,k>0 \\ k=1,\dots i-1}}&=-s_yC_{k,i-k}\overbrace{+2s_r\sum_{h=\big[\frac{i}{2}\big]}^{\infty} \binom{i}{k} p_r^{k}(1-p_r)^{i-k}C_{0,h} \binom{2h}{i}\frac1{2^{2h}}}^{\text{Pure red $\longrightarrow$ mix ($k,i-k$)}}\\
    &\overbrace{+2s_y\sum_{\substack{j+h=\big[\frac{i}{2}  \big]\\j>0,h=\big[\frac{i-k}{2}\big]}}^{\infty} \sum_{v=0}^{k} \binom{i-k}{k-v} p_r^{k-v}(1-p_r)^{i-(k-v)}C_{j,h} \binom{2j}{v}\binom{2h}{i-v}\frac1{2^{2(j+h)}}}^{\text{Mix $\longrightarrow$ mix ($k,i-k$)}}.\\
\end{aligned}
$}
\end{equation}

Finally, combining (\ref{yellow}), (\ref{red}) and (\ref{mix}) (and (\ref{mixoneway1}) or (\ref{mixoneway2}) in case of one-way switching)  together with the dynamics for cells carrying $0$ ecDNA copies at time $t$, namely $C_{0,0}(t)$, we get the complete system of equations, which represents a master equation in discrete time. In case of two-way switching, i.e. $p_y, p_r>0$:
\begin{equation}\label{sistemaswitch} 
\resizebox{.95\hsize}{!}{$
\begin{aligned}
    &\frac{dC_{i,0}(t)}{dt}\bigg|_{i>0}=-s_yC_{i,0}+2s_y\sum_{j=\big[\frac{i}{2}  \big]}^{\infty} (1-p_y)^i C_{j,0} \binom{2j}{i}\frac1{2^{2j}}\\
    &+2s_r\sum_{h=\big[\frac{i}{2}  \big]}^{\infty} p_r^i C_{0,h} \binom{2h}{i}\frac1{2^{2h}}+2s_y\sum_{\substack{j+h=\big[\frac{i}{2}\big]\\j, h>0 }}^{\infty} (1-p_y)^i C_{j,h} \binom{2j}{i}\binom{2h}{0}\frac1{2^{2(j+h)}}\\
    &+2s_y\sum_{\substack{j+h=\big[\frac{i}{2}\big]\\j, h>0 }}^{\infty}\sum_{v=0}^{i-1} \binom{i}{v} p_r^{i-v} (1-p_y)^{v}C_{j,h} \binom{2j}{v}\binom{2h}{i-v}\frac1{2^{2(j+h)}}\\
    &\frac{dC_{0,i}(t)}{dt}\bigg|_{i>0}=-s_rC_{0,i}+2s_y\sum_{j=\big[\frac{i}{2}  \big]}^{\infty} p_y^i C_{j,0} \binom{2j}{i}\frac1{2^{2j}}\\
    &+2s_r\sum_{h=\big[\frac{i}{2}  \big]}^{\infty} (1-p_r)^i C_{0,h} \binom{2h}{i}\frac1{2^{2h}}+2s_y\sum_{\substack{j+h=\big[\frac{i}{2}\big]\\j, h>0 }}^{\infty} (1-p_r)^i C_{j,h} \binom{2j}{0}\binom{2h}{i}\frac1{2^{2(j+h)}}\\
    &+2s_y\sum_{\substack{j+h=\big[\frac{i}{2}\big]\\j, h>0 }}^{\infty}\sum_{v=1}^{i} \binom{i}{v} p_y^{v} (1-p_r)^{i-v}C_{j,h} \binom{2j}{v}\binom{2h}{i-v}\frac1{2^{2(j+h)}}\\
    &\frac{dC_{k,i-k}(t)}{dt}\bigg|_{\substack{i,k>0 \\ k=1,\dots i-1}}=-s_yC_{k,i-k}+2s_y\sum_{j=\big[\frac{i}{2}\big]}^{\infty} \binom{i}{k} p_y^{i-k}(1-p_y)^{k}C_{j,0} \binom{2j}{i}\frac1{2^{2j}}\\
    &+2s_r\sum_{h=\big[\frac{i}{2}\big]}^{\infty}p_r^{k}(1-p_r)^{i-k}C_{0,h} \binom{2h}{i}\frac1{2^{2h}}\\
    &+2s_y\sum_{\substack{j+h=\big[\frac{i}{2}  \big]\\j,h>0}}^{\infty}\sum_{v=1}^{i}\sum_{m=1}^{k} \binom{i-v}{k-m} \binom{v}{m} p_y^{v-m}(1-p_y)^{m}p_r^{k-m}(1-p_r)^{i-v-(k-m)}C_{j,h} \binom{2j}{v}\binom{2h}{i-v}\frac1{2^{2(j+h)}}\\
    &\frac{dC_{0,0}}{dt}=-C_{0,0}+2C_{0,0}+2s_y\sum_{\substack{j+h=1\\j>0}}^\infty C_{j,h}\binom{2(j+h)}{0}\frac1{2^{2(j+h)}}+2s_r\sum_{h=0}^\infty C_{0,h}\binom{2h}{0}\frac1{2^{2h}}.\\
\end{aligned}
$}
\end{equation}
Similarly, we can obtain the system for one-way switching by combining equations (\ref{yellow}) and (\ref{red}) with either equation (\ref{mixoneway1}) or (\ref{mixoneway2}), based on the values of $p_y$ and $p_r$.

\begin{table}[H]
    \centering
    \resizebox{\columnwidth}{!}{%
\begin{tabular}{|c|c|c|}
    $p_y$ &probability of switching from yellow to red ecDNA\\
    $p_r$ &probability of switching from red to yellow ecDNA\\
    $s_y$ &reproduction rate for cells carrying only yellow ecDNA\\
    $s_r$ &reproduction rate for cells carrying only red ecDNA\\
    $n^+(t)$ &proportion of cells carrying at least one ecDNA copy\\
    $n^-(t)$ &proportion of cells carrying no ecDNA copies\\
    $\gamma(t)$ &loss rate of ecDNA due to a completely asymmetric segregation\\
    $C_{i,k}(t)$ &number of cells carrying $i$ yellow and $k$ red ecDNA copies at time $t$\\
    $\rho_{i,k}(t)$ &frequency of cells carrying $i$ yellow and $k$ red ecDNA copies at time $t$\\
    $g_{1,2}(t), h_{1,2}(t), k_{1,2}(t)$ &loss/gaining rates for ecDNA evolutionary process\\
    $M(t)$ &population growth factor at time $t$\\
    $N_k(t)$ &number of cells carrying $k$ ecDNA copies at time $t$\\
    $\delta(s)$ &zeroing function for ecDNA-free cells trend\\
    $\Delta(s)$ &normalising function for ecDNA-free cells trend\\
    $\mathbf{M}^{(l)}(t)$ &$l$-order moment for the whole population at time $t$ \\
    $\mathbf{M}^{(l)}_y(t)$ &$l$-order moment for pure yellow ecDNA cells at time $t$ \\
    $\mathbf{M}^{(l)}_r(t)$ &$l$-order moment for pure red ecDNA cells at time $t$ \\
    $\mathbf{M}^{(l)}_m(t)$ &$l$-order moment for mix ecDNA cells at time $t$\\
\end{tabular}%
}
\caption[Parameters and variables in the mathematical description of multiple ecDNA types]{Notations for the mathematical description of multiple ecDNA types}
    \label{tab1}
\end{table}

\section*{Results}

We are interested in the dynamics of pure, mix and ecDNA-free cells. Thus, we investigate how selection strength and switching parameters lead to different patterns. Again in the results, $p_y=p_r=0$ refers to multiple ecDNA species, otherwise it refers to multiple ecDNA genotypes or phenotypes. $s_y=s_r=1$ indicates neutral selection, $s_y=s_r>1$ indicates identical positive selection, and $s_y\neq s_r\geq1$ is for non-identical positive selection.

\subsection*{Copy number distribution of all ecDNA-positive cells and fraction of ecDNA-free cells over time}

\subsubsection*{Under neutral and identical positive selection}

Let us first look at the scenarios where both yellow and red ecDNAs have the same reproduction rate, i.e. $s_y=s_r=s$, where $s\ge1$. From Eqs. \ref{sistemaswitch}, setting $s_y=s_r=s$ as stated, we can recover the dynamics of number of cells with $k$ ecDNA copies at time $t$, $N_k(t)$ as
\textcolor{black}{
\begin{equation}\label{eqN1}
\begin{aligned}
    \frac{dN_k(t)}{dt}&=\frac{d\big(C_{k,0}(t)+C_{0,k}(t)+\sum_{k=1}^{i-1}C_{k,i-k}\big|_{i,k>0}(t)\big)}{dt}\\
    &=-sN_k(t)+2s\sum_{i=\big[\frac{k}{2}  \big]}^{\infty} N_i(t)\binom{2i}{k}\frac1{2^{2i}} \,.
    \end{aligned}
\end{equation}
}
\textcolor{black}{
This can be obtained by summing the right hand sides of the first three equations in (\ref{sistemaswitch}). A detailed breakdown follows.\\
We first sum the terms related to the contributions to each compartment of pure yellow's daughters, i.e. the sums containing the terms in $C_{j,0}$. These sums to:
\begin{equation}\label{equationsum1draft}
    \sum_{j=\big[\frac{i}2\big]}^\infty\sum_{k=0}^i \binom{i}{k} p_y^{i-k}(1-p_y)^{k} C_{j,0} \binom{2j}{i} \frac1{2^{2j}}.
\end{equation}
We notice that the term $\sum_{k=0}^i \binom{i}{k} p_y^{i-k}(1-p_y)^{k}=1$, as it represents the total probability of all possible outcomes in a sequence of $i$ independent Bernoulli trials, where each trial has probability $p_y$ of success. Thus, this sum is the total probability space of all the events in a Binomial distribution. We can conclude that (\ref{equationsum1draft}) is the sum of all the possible events related to the birth of a daughter with $i$ copies from a pure yellow mother, thus:
\begin{equation}\label{equationsum1}
    \sum_{j=\big[\frac{i}2\big]}^\infty\sum_{k=0}^i \binom{i}{k} p_y^{i-k}(1-p_y)^{k} C_{j,0} \binom{2j}{i}\frac1{2^{2j}}=\sum_{j=\big[\frac{i}2\big]}^\infty C_{j,0} \binom{2j}{i} \frac1{2^{2j}}.
\end{equation}
Similarly, the sum of the terms related to the contributions to each compartment of pure red's daughters, i.e. the sums containing the terms in $C_{0,h}$, gives:
\begin{equation}\label{equationsum2}
    \sum_{h=\big[\frac{i}2\big]}^\infty\sum_{k=0}^i \binom{i}{k} p_r^{k}(1-p_r)^{i-k} C_{j,0} \binom{2h}{i}\frac1{2^{2h}}=\sum_{h=\big[\frac{i}2\big]}^\infty C_{0,h} \binom{2h}{i} \frac1{2^{2h}}.
\end{equation}
Finally, the sum of the terms related to the contributions to each compartment of mix's daughters, i.e. the sums containing the terms in $C_{j,h}$, is:
\begin{equation}\label{equationsum3draft}
\resizebox{.99\hsize}{!}{$
\begin{aligned}
    \sum_{\substack{j+h=\big[\frac{i}2\big]\\ j,h>0}}^\infty\sum_{k=0}^i\sum_{v=0}^i \sum_{m=0}^k \binom{i-v}{k-m} \binom{v}{m} p_y^{v-m}(1-p_y)^{m}p_r^{k-m}(1-p_r)^{i-v-(k-m)} C_{j,h} \binom{2j}{v} \binom{2h}{i-v} \frac1{2^{2j+2h}}.
\end{aligned}
$}
\end{equation}
The above term is more complex than the previous sums. However, we can notice that the term $\sum_{k=0}^i\sum_{v=0}^i \sum_{m=0}^k \binom{i-v}{k-m} \binom{v}{m} p_y^{v-m}(1-p_y)^{m}p_r^{k-m}(1-p_r)^{i-v-(k-m)}=1$ as it represents the sum of all possible outcomes in a convolution of two Binomial distributions (one with success probability $p_y$ and another with success probability $p_r$). We can then notice that the term in (\ref{equationsum3draft}) is the sum of all possible events relative to the birth of a daughter with $i$ copies from a mix mother. Thus we can claim that:
\begin{equation}\label{equationsum3}
\resizebox{.95\hsize}{!}{$
\begin{aligned}
    &\sum_{\substack{j+h=\big[\frac{i}2\big]\\ j,h>0}}^\infty\sum_{k=0}^i\sum_{v=0}^i \sum_{m=0}^k \binom{i-v}{k-m} \binom{v}{m} p_y^{v-m}(1-p_y)^{m}p_r^{k-m}(1-p_r)^{i-v-(k-m)} C_{j,h} \binom{2j}{v} \binom{2h}{i-v} \frac1{2^{2j+2h}}\\
    &=\sum_{\substack{j+h=\big[\frac{i}2\big]\\ j,h>0}}^\infty C_{j,h} \binom{2j+2h}{i} \frac1{2^{2j+2h}}.
\end{aligned}
$}
\end{equation}
By summing the three terms in (\ref{equationsum1}), (\ref{equationsum2}) and (\ref{equationsum3}), together with the negative terms relative to the division of a pure or mix mother, we get the equality in (\ref{eqN1}).\\}
Furthermore, we can extend (\ref{eqN1}) to the identical positive selection case:
\begin{equation}\label{eqN2}
    \begin{aligned}
        \frac{dN_k(t)}{dt}\bigg|_{k>0}&=s\bigg[\frac{dN_k(t)}{dt}\bigg|_{s=1}    \bigg],\\
        \frac{dN_0(t)}{dt}&=s\bigg[\frac{dN_0(t)}{dt}\bigg|_{s=1}    \bigg]-(s-1)N_0.
    \end{aligned}
\end{equation}
\textcolor{black}{
This consists in writing the system as a branching process of a single-ecDNA type and it is in line withe the modeling proposed in \cite{22}.
}
Using the definition of cell frequency (see Table \ref{tab1}):
\begin{equation}\label{eqdetdem}
    \rho_{i,k}(t)=\frac{C_{i,k}(t)}{M(t)}
\end{equation}
together with Eqs. (\ref{eqN1}) and (\ref{eqN2}), we can find that the total population size, $M(t)$ under different selection scenarios as
\begin{equation}\label{eqM}
    \begin{aligned}
        &M(t)=e^t, \qquad &\text{for $s=1$,}\\
        &M(t)=e^{st-(s-1)\int_0^t\rho_0(\tau)\,d\tau},\qquad &\text{for $s>1$}.
    \end{aligned}
\end{equation}
Eqs. (\ref{eqN2}) and (\ref{eqM}) show that, when the selection strength is identical for cells carrying any of the two ecDNA types, the distribution of total ecDNA copies remains to be independent of the switching rates over time (Figure \ref{F:invariant}c).

This independence from switching probabilities holds for the fraction of ecDNA-free cells over time as well (Figure \ref{F:invariant}a). Indeed, using Eqs. (\ref{sistemaweini}) based on our methods of modeling single ecDNA types in \cite{20}, we can obtain solutions for $n^-(t)$ and get the frequency of ecDNA-free cells as
\begin{equation}\label{rho00}
\begin{aligned}
    &\rho_{0,0}(t)=\frac{t}{2+t}, \qquad \text{if $s_y=s_r=1$},\\
    &\text{and} \\
    &\rho_{0,0}(t)=\frac{t}{2+te^{\delta(s)t}}, \qquad \text{if $s_y=s_r>1$}.
\end{aligned}
\end{equation}
Whilst the fraction of ecDNA free cells, $\rho_{0,0}$, is approaching 1 as $t\to\infty$ for neutral selection ($s_y=s_r=1$), we expect the progressive decrease of the fraction of cells carrying ecDNA. On the contrary, $\rho_{0,0}$ shrinks to zero under identical positive selection ($s_y=s_r>1$). 


\subsubsection*{Under non-identical fitness of cells carrying different types of ecDNAs}

\begin{figure}[H]
	\centering
	\includegraphics[width=1\textwidth]{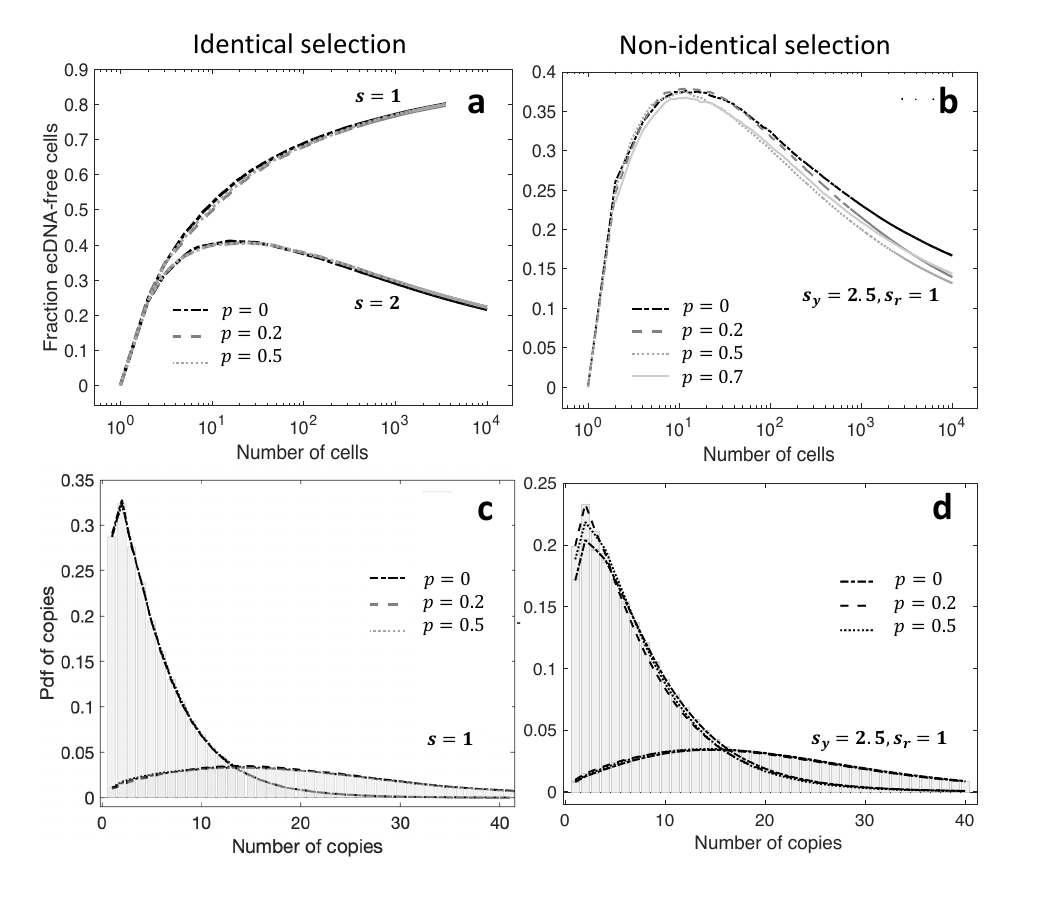}
	\caption[ecDNA copy distribution among either identical or non-identical fitness scenarios]{\textbf{ecDNA copy distribution among either identical or non-identical selection.} Under identical (neutral and positive) selection for both ecDNA types \textbf{(a)}, the fraction of ecDNA-free cells over time does not change with switching probabilities. Whilst if non-identical fitness is considered \textbf{(b)}, the fraction of ecDNA-free cells will depend on the probabilities of type switching. These panels a and b are realised starting from a single cell with 1 yellow copies and the trends are recorded over 1000 realisations. 
    Similarly, the copy number distribution of all cells carrying ecDNA is independent on $p$ when identical fitness scenario is considered \textbf {(c)}, whilst it changes with the rate of type switching if yellow and red ecDNA carry different reproduction rates \textbf{(d)}. For the copy number distributions (c and d), we showed outcomes from two different initial conditions, for the highest curves we started with a single cell with 1 yellow and 1 red ecDNA copy, and for the lowest curves we had a starting single cell with 10 yellow and 10 red ecDNA copies ($p_y=p_r=p$). Distributions are recorded at $10^4$ cells over 1000 realisations.}
	\label{F:invariant}
\end{figure}

Now we look at the scenario where fitness for cells carrying different ecDNAs is not identical ($s_y\neq s_r$). The copy number distribution of total ecDNA-positive cells and the fraction of ecDNA-free cells are highly sensitive to selection coefficients ( Figure \ref{F:invariant}b\&d). The independence of the those quantities on the switching rates, $p_y$ and $p_r$, is lost.  

\subsection*{Moment analysis for multiple ecDNA species when switching is off}

We can study the dynamics of two ecDNA species by plugging $p_y=p_r=0$ in Eqs. (\ref{sistemaswitch}), obtaining the following demographic system as

\begin{equation}\label{sistemaswitchspecies}
\resizebox{.95\hsize}{!}{$
\begin{aligned}
    \frac{dC_{i,0}(t)}{dt}\bigg|_{i>0}&=-s_yC_{i,0}+2s_y\sum_{j=\big[\frac{i}{2}  \big]}^{\infty} C_{j,0} \binom{2j}{i}\frac1{2^{2j}}+2s_y\sum_{j=[\frac{i}{2}],h=1}^{\infty} C_{j,h}\binom{2j}{i}\binom{2h}{0}\frac1{2^{2(j+h)}},\\
    \frac{dC_{0,i}(t)}{dt}\bigg|_{i>0}&=-s_rC_{0,i}+2s_r\sum_{h=\big[\frac{i}{2}  \big]}^{\infty} C_{0,h} \binom{2h}{i}\frac1{2^{2h}}+2s_y\sum_{j=1,h=[\frac{i}{2}]}^{\infty} C_{j,h}\binom{2j}{0}\binom{2h}{i}\frac1{2^{2(j+h)}},\\
    \frac{dC_{k,i-k}(t)}{dt}\bigg|_{i,k>0}&=-s_yC_{k,i-k}(t)+2s_y\sum_{j=\big[\frac{k}{2}  \big],\,h=\big[\frac{i-k}{2}  \big]}^{\infty} C_{j,h}\binom{2j}{k}\binom{2h}{i-k}\frac1{2^{2(j+h)}}\\
    \frac{dC_{0,0}}{dt}&=-C_{0,0}+2C_{0,0}+2s_y\sum_{\substack{i+k=1\\i>0}}^\infty C_{i,k}\binom{2(i+k)}{0}\frac1{2^{2(i+k)}}+2s_r\sum_{k=0}^\infty C_{0,k}\binom{2k}{0}\frac1{2^{2k}}.\\
\end{aligned}
$}
\end{equation}


Using Eqs. (\ref{eqM}) and (\ref{sistemaswitch}), we calculate the $l$-order moment for each subpopulation by the formula
\begin{equation}\label{moment}
    \mathbf{M}_j^{(l)}(t)=\sum_{i,k}(i+k)^l\rho_{i,k},
\end{equation}
where $j=y, r, m$ is respectively for pure yellow, pure red and mix subpopulations. \\
We focus on the case of identical fitness, i.e. $s_y=s_r=s$ including scenarios under neutral selection $s=1$ and positive selection $s>1$. Starting with $s=1$, we write exact equations for probability densities of pure and mix cells over time using Eqs. (\ref{eqdetdem}). For example, for pure yellow cells, we have

\begin{equation}\label{densbluespecies}
    \frac{d\rho_{i,0}(t)}{dt}=-2\rho_{i,0}(t)+2\sum_{j=[\frac{i}{2}]}^{\infty}\rho_{j,0}(t)\binom{2j}{i}\frac1{2^{2j}}+2\sum_{j=[\frac{i}{2}],h=1}^{\infty} \rho_{j,h}(t)\binom{2j}{i}\frac1{2^{2(j+h)}}
\end{equation}
Then, Eq. (\ref{densbluespecies}) can be substituted in Eq. (\ref{moment}) to get every $l$-order moment. The first moment dynamic, which describes the average copy number of pure cells carrying only yellow ecDNA copies, is given by
\begin{equation}\label{metodomom}
\resizebox{.95\hsize}{!}{$
\begin{aligned}
    \frac{d\mathbf{M_y}^{(1)}(t)}{dt}&=\sum_{i=0}^{\infty}i\,d\rho_{i,0}(t)\\
    &=-2\mathbf{M_y}^{(1)}(t)+2\sum_{i=0}^{\infty}\sum_{j=\big[\frac{i}{2}  \big]}^{\infty}i\rho_{j,0}(t)\binom{2j}{i}\frac1{2^{2j}}\\
    & \quad +\sum_{i=0}^{\infty}\sum_{j=\big[\frac{i}{2}  \big], h=1}^{\infty}i\rho_{j,h}(t)\binom{2j}{i}\frac1{2^{2(j+h)}}\\
    &=-2\mathbf{M_y}^{(1)}(t)+2\sum_{j=0}^{\infty}\rho_{j,0}(t)\frac1{2^{2j}}\sum_{i=0}^{2j}i\binom{2j}{i}+2\sum_{j,h=1}^{\infty}\rho_{j,h}(t)\frac1{2^{2(j+h)}}\sum_{i=0}^{2j}i\binom{2j}{i}\\
    &=-2\mathbf{M_y}^{(1)}(t)+2\sum_{j,h=0}^{\infty}\rho_{j,0}(t)\frac1{2^{2j}}2^{2j}j+2\sum_{j,h=1}^{\infty}\rho_{j,h}(t)\frac1{2^{2(j+h)}}2^{2j}j\\
    &=-2\mathbf{M_y}^{(1)}(t)+2\mathbf{M_y}^{(1)}(t)+2\sum_{j,h=1}^{\infty}j\rho_{j,h}(t)\frac1{2^{2h}}=2\mathbf{E}_j^\infty(\rho_{j,h}).
    \end{aligned}
    $}
\end{equation}
where 
\begin{equation}\label{ej}
\mathbf{E}_j^\infty(\rho_{j,h})=\sum_{j,h=1}^{\infty}j\rho_{j,h}(t)\frac1{2^{2h}}
\end{equation}
is the infinite sum above, representing the expected value of $\rho$ weighted on $j$ and divided by a geometric term in $h$.
We notice that $\mathbf{E}_j^\infty(\rho_{j,h})$ is a sum whose general term decays geometrically with $h$. We then expect that the terms of this sum are getting progressively smaller as $h\rightarrow\infty$. We can then approximate the above equation with the first 4 terms of the sum, namely $\mathbf{E}_j^4(\rho_{j,h})=\frac14\sum_{j=1}^{\infty}j\rho_{j,1}(t)+\frac1{16}\sum_{j=1}^{\infty}j\rho_{j,2}(t)+\frac1{64}\sum_{j=1}^{\infty}j\rho_{j,3}(t)+\frac1{256}\sum_{j=1}^{\infty}j\rho_{j,4}(t)$, getting then:
\begin{equation}\label{e4}
    \frac{d\mathbf{M_y}^{(1)}(t)}{dt}\approx2\mathbf{E}_j^4(\rho_{j,h})
\end{equation}
Similarly, we can obtain equations for the pure red and mix subpopulations as in Table (\ref{tabmomapprox1species}, by also defining:
\[
\mathbf{E}_h^\infty(\rho_{j,h})=\sum_{j,h=1}^{\infty}h\rho_{j,h}(t)\frac1{2^{2j}}
\]

Now we move to the scenario of identical positive fitness, i.e $s_y=s_r=s>1$, combining Eqs. (\ref{eqN2}) and (\ref{eqM}), we obtain the frequencies of pure yellow ecDNA cells as
\begin{equation}\label{densbluespeciessel}
    \frac{d\rho_{i,0}(t)}{dt}=s\bigg[\frac{d\rho_{i,0}(t)}{dt}\bigg|_{s=1}    \bigg]-(s-1)\rho_{0,0}\rho_{i,0}.
\end{equation}
Using the moment calculation formula in (\ref{moment}) we get
\begin{equation}\label{densblu2}
\begin{aligned}
    \frac{d\mathbf{M_y}^{(1)}(t)}{dt}&=s\bigg[\frac{d\mathbf{M}_y^{(1)}(t)}{dt}\bigg|_{s=1} \bigg]-(s-1)\mathbf{M_y}^{(1)}(t)\rho_{0,0}\\
    &=2s\mathbf{E}_j^\infty(\rho_{j,h}(t))-(s-1)\mathbf{M_y}^{(1)}(t)\rho_{0,0}\\
    &\approx2s\mathbf{E}_j^4(\rho_{j,h}(t))-(s-1)\mathbf{M_y}^{(1)}(t)\rho_{0,0}.
\end{aligned}
\end{equation}
The same steps in (\ref{densbluespecies}) and (\ref{densblu2}), together with standard solution techniques for ODEs, are used for filling Table \ref{tabmomapprox1species}. The total moment $\mathbf{M}^{(1)}$ has been already given \cite{22}:
\begin{equation}\label{totmom}
    \begin{aligned}
    &\mathbf{M}^{(1)}(t)=1, \qquad &\text{if $s=1$},\\
    &\text{and}\\
    &\mathbf{M}^{(1)}(t)=e^{(s-1)\int_0^t\rho_{0,0}(\tau)\,d\tau},\qquad &\text{if $s>1$}.
    \end{aligned}
\end{equation}

The results for the first moment dynamics in the neutral case, as shown in Equation (\ref{totmom}), are also consistent with those obtained in models of other extra-chromosomal elements that undergo binomial partitioning and duplication during cell division, such as mitochondrial DNA, as described in the models by Johnston et al. \cite{Jones1, Jones2}. In those studies, the dynamics of mtDNA include replication occurring outside of cell division. However, when replication outside of division is excluded, their model simplifies to the framework presented here, and the analytical results for a single-type evolutionary process align with ours.

\begin{table}[H]
    \centering
    \resizebox{\columnwidth}{!}{%
    \begin{tabular}{ |c|c|c|  }
    \hline
    \multicolumn{3}{|c|}{\textbf{Weighted first moment dynamics equations for ecDNA species}}\\
    \hline
    \multicolumn{3}{|c|}{Neutral case $s_y=s_r=s=1$} \\
    \hline
    Subpopulation & & Equation \\
    \hline
    Total &$\frac{d\mathbf{M}^{(1)}(t)}{dt}=$  &$0$ \\
    Pure yellow &$\frac{d\mathbf{M_y}^{(1)}(t)}{dt}=$ &$2\mathbf{E}_j^4(\rho_{j,h})(t)$\\
    Pure red &$\frac{d\mathbf{M_r}^{(1)}(t)}{dt}=$ &$2\mathbf{E}_h^4(\rho_{j,h})$ \\
    Mix &$\frac{d\mathbf{M_m}^{(1)}(t)}{dt}=$ &$-2\mathbf{E}_j^4(\rho_{j,h})-2\mathbf{E}_h^4(\rho_{j,h})$ \\
    \hline
    \multicolumn{3}{|c|}{Identical positive selection case $s_y=s_r=s>1$} \\
    \hline
    Subpopulation & & Equation \\
    \hline
    Total &$\frac{d\mathbf{M}^{(1)}(t)}{dt}=$  &$(s-1)\rho_{0,0}\mathbf{M}^{(1)}(t)$ \\
    Pure yellow &$\frac{d\mathbf{M_y}^{(1)}(t)}{dt}=$ &$2s\mathbf{E}_j^4(\rho_{j,h}(t))-(s-1)\mathbf{M_y}^{(1)}(t)\rho_{0,0}$ \\
    Pure red &$\frac{d\mathbf{M_r}^{(1)}(t)}{dt}=$ &$2s\mathbf{E}_h^4(\rho_{j,h}(t))-(s-1)\mathbf{M_r}^{(1)}(t)\rho_{0,0}$ \\
    Mix &$\frac{d\mathbf{M_m}^{(1)}(t)}{dt}=$ &by subtraction \\ 
    \hline
\end{tabular}%
}
\caption[Equations for weighted first moment dynamics, ecDNA species]{Summary of equations for weighted first moment dynamics for ecDNA species in the identical fitness coefficients cases.}
    \label{tabmomapprox1species}
\end{table}

In this and the following sections, we compare analytical solutions with simulations where possible. The deviation between simulations and analytical results will be quantified using the mean squared error (MSE):
\begin{equation}
    \text{MSE}=\frac1n\sum_{i=1}^n(y_i-\hat{y}_i)^2,
\end{equation}
where $n$ is the number of observations, $y_i$ is the expected value for each observation, i.e. the analytical solution, and $\hat{y}_i$ is the simulated value.\\
Additionally, we will present 95\% confidence intervals for the mean results obtained from multiple realisations. These intervals will be shown for all weighted moment dynamics and absolute mean copy number trends\cite{Cumming2007}.\\
Finally, one time unit corresponds to a single cell division. As a result, time is measured as number of cells over time. The $x$-axis will be then labeled as $2^{i}$ cells, with $i=1,2,3,\dots$.

\begin{figure}[H]
	\centering
	\includegraphics[width=1\textwidth]{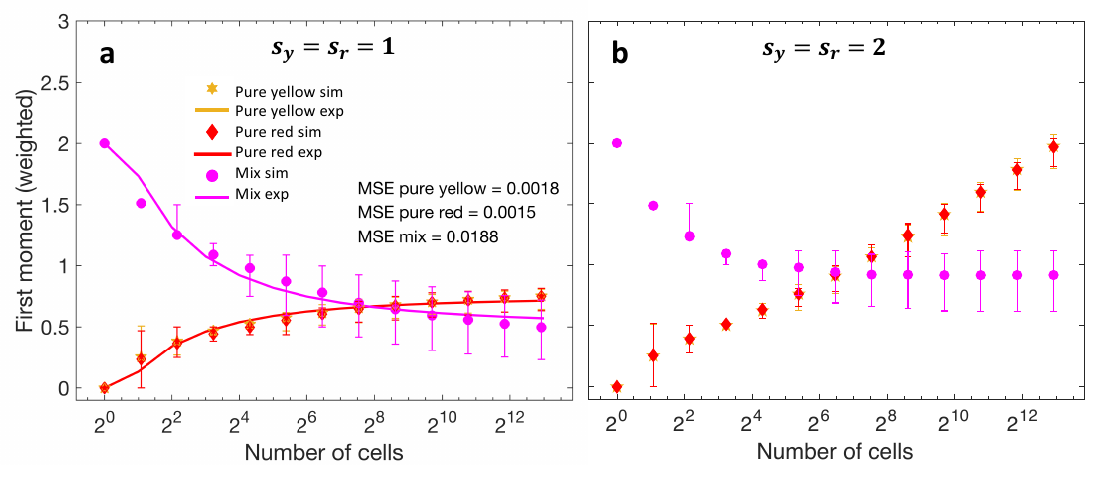}
	\caption[Weighted first moment dynamics for ecDNA species]{\textbf{Weighted first moment dynamics for ecDNA species ($p_y=p_r=p=0$).} This refers to the contribution of pure and mix cells to the mean copy number of the total population. We compare simulations data with analytical solutions for first moment dynamics of ecDNA species. \textbf{a} shows the \textbf{neutral selection} case, whilst \textbf{b} refers to \textbf{identical positive selection}. Solid lines represent analytical solutions, whilst scatter dots refer to simulations, and mean squared errors (MSE) are displayed as a measure for the distance between simulations and analytics. In both panels, 95\% confidence intervals for the mean first moments are shown. The initial condition is a single cell with 1 yellow and 1 red ecDNA copy, and this is why the trends for first pure yellow and pure red ecDNA moments are coinciding. Results are averaged over 1000 realisations.}
	\label{F:figuraspecies}
\end{figure}

Our analytical results for the neutral case, obtained by solving numerically the equations of Table \ref{tabmomapprox1species}, match well with stochastic simulations implemented by Gillespie Algorithm \cite{32,33,34,35} (Figure \ref{F:figuraspecies}a). 
Our moment analysis of each sub-populations refers to weighted moments, which considers the variability in copy numbers across different subpopulations together with their respective relative fractions to the total population size. Specifically, the weighted mean copy number are normalised based on the mean copy number for the total population. For example, the mean copy number for the total population used for the normalisation is constant under neutral selection, which is $2$ as the copy number of the initial cell in Figure \ref{F:figuraspecies}a. Under identical positive selection, it is $2t$ in Figure \ref{F:figuraspecies}b. 

The dynamics of weighted moments align with the trends observed in the unweighted mean copy number dynamics of each subpopulation, as illustrated in Figure \ref{F:meancopynumberspecies}. Here, we examine every subpopulation separately and tracks their absolute mean copy numbers over time. As expected, the absolute average copy numbers increases over time for ecDNA-positive cells no matter they are pure or mix cells. 
However, the changes in curve slopes portrayed by the weighted moments in Figure \ref{F:figuraspecies} suggest that, despite the steady increase in the absolute number of ecDNA copies in both pure and mix cells, the proportion of pure cells contributes more substantially than mix ones to the mean copy number of the total population if selection are identical (neutral and positive).

While it is difficult to obtain analytical solutions for the non-identical selection case, we simulate the weighted first moment (Figure \ref{F:figuraspeciesnonid}a) and absolute mean copy number (Figure \ref{F:figuraspeciesnonid}b) for the case of two ecDNA species, where the cells carrying yellow ecDNA are assumed to proliferate faster than the cells carrying red ecDNA. Conversely to the dynamics under identical positive selection (Figure \ref{F:figuraspecies}b and Figure \ref{F:meancopynumberspecies}b), pure yellow ecDNA cells dominate the populations over time as expected (Figure \ref{F:figuraspeciesnonid}a). However, since we assume that mix cells have the same fitness as the fitter pure cells, the contribution of mix cells to the whole population is much less than the one of pure cells, no matter if under identical positive selection or non-identical selection. In the infinite time limit, it seems that the mix cells cannot be maintained.

Across all fitness scenarios (Figures \ref{F:figuraspecies}-\ref{F:figuraspeciesnonid}) , the largest fluctuations, although still relatively small, are observed in the dynamics of mix cells, highlighting a higher copy number variability within this subpopulation. This is expected, as the absence of switching prevents pure cells from becoming mix. As a result, the mix subpopulation exhibits the most variability among all the subpopulations, since mix cells can produce daughters in all possible compartments (i.e., mix, pure, and ecDNA-free) due to binomial segregation during division.

\subsection*{Moment analysis of multiple ecDNA genotypes or phenotypes when switching is on}

\subsubsection*{Under identical fitness and one-way switching}

Now we analyse the dynamics of two ecDNA genotypes or phenotypes, where different types of ecDNA elements can switch to each other. We first look at a simpler case of one-way switching. Without the loss of generality, we can set $p_r=0$ and $p_y=p$.
We consider values for which $p\ll1$, whose are more biologically relevant. 
Using the Taylor expansion, we can write
\begin{equation}\label{approximationfactors}
    \begin{aligned}
        &1-ip<(1-p)^i<1-p, \qquad 0<p^i<p, \qquad p\ll1;\\
    \end{aligned}
\end{equation}
We use the notation $p_l$ and $p_u$ to indicate the lower-bound and upper-bound approximations for $p$ in (\ref{approximationfactors}).

In the simplest case of neutral selection, i.e. $s_y=s_r=s=1$, we write exact equations for frequencies of pure, mix and ecDNA-free cells over time using Eqs. (\ref{eqdetdem}). For the pure yellow subpopulation, we have
\textcolor{black}{
\begin{equation}\label{densblue}
\begin{aligned}
    \frac{d\rho_{i,0}(t)}{dt}=&-2\rho_{i,0}(t)+2\sum_{j=[\frac{i}{2}]}^{\infty}(1-p_y)^i\rho_{j,0}(t)\binom{2j}{i}\frac1{2^{2j}}+2\sum_{\substack{j+h=[\frac{i}{2}]\\j,h>0}}^{\infty}(1-p_y)^i\rho_{j,0}(t)\binom{2j}{i}\frac1{2^{2j}}\\.
\end{aligned}
\end{equation}
}
Then, we can substitute Eq. (\ref{densblue}) in (\ref{moment}) to get every $l$-order moment. The first moment dynamic, i.e. the average copy number of cells carrying only yellow ecDNA elements, using the \textcolor{black}{$p_u$} approximation:
\textcolor{black}{
\begin{equation}\label{metodomomoneway}
\resizebox{.98\hsize}{!}{$
\begin{aligned}
    \frac{d\mathbf{M_y}^{(1)}(t)}{dt}&=\sum_{i=0}^{\infty}i\,d\rho_{i,0}(t)\\
    &=-2\mathbf{M_y}^{(1)}(t)+2\sum_{i=0}^\infty i\sum_{j=[\frac{i}{2}]}^{\infty}(1-p)^i\rho_{j,0}(t)\binom{2j}{i}\frac1{2^{2j}}+2\sum_{i=0}^\infty i \sum_{\substack{j+h=[\frac{i}{2}]\\j,h>0}}^{\infty}(1-p)^i\rho_{j,0}(t)\binom{2j}{i}\frac1{2^{2j}}\\
    &=-2\mathbf{M_y}^{(1)}(t)+2\sum_{j=0}^\infty \rho_{j,0}(t)\frac1{2^{2j}} \sum_{i=0}^{2j}i(1-p)^i\binom{2j}{i}+2\sum_{j,h=1}^\infty \rho_{j,h}(t)\frac1{2^{2(j+h)}} \sum_{i=0}^{2j}i(1-p)^i\binom{2j}{i}\\
    &\approx-2\mathbf{M_y}^{(1)}(t)+2(1-p)\sum_{j=0}^\infty \rho_{j,0}(t)\frac1{2^{2j}} \sum_{i=0}^{2j}i\binom{2j}{i}+2(1-p)\sum_{j,h=1}^\infty \rho_{j,h}(t)\frac1{2^{2(j+h)}} \sum_{i=0}^{2j}i\binom{2j}{i}\\
    &=-2\mathbf{M_y}^{(1)}(t)+2(1-p)\mathbf{M_y}^{(1)}(t)+2(1-p)\mathbf{E}_j^\infty(\rho_{j,h})\\
    &\approx -2p\mathbf{M_y}^{(1)}(t)+2(1-p)\mathbf{E}_j^4(\rho_{j,h})
  \end{aligned}
  $}
\end{equation}
The approximation $(1-p)^i\approx 1-p$ has been applied. The last term can be obtained as in the previous section (Eq. (\ref{metodomom}) and it is $\mathbf{E_j^\infty}(\rho_{j,h})$ (Eq. (\ref{ej})). This can be approximated with the sum of the first 4 terms, like in (\ref{e4}).\\ 
Using the approximation with $p_l$, i.e. $(1-p)^i\approx 1-ip$, $p^i\approx 0$, we get similarly:
\begin{equation}
\resizebox{.98\hsize}{!}{$
\begin{aligned}
    \frac{d\mathbf{M_y}^{(1)}(t)}{dt}&\approx -2\mathbf{M_y}^{(1)}(t)+2\sum_{j=0}^\infty \rho_{j,0}(t)\frac1{2^{2j}} \sum_{i=0}^{2j}i(1-ip)\binom{2j}{i}+2\sum_{j,h=1}^\infty \rho_{j,h}(t)\frac1{2^{2(j+h)}} \sum_{i=0}^{2j}i(1-ip)\binom{2j}{i}\\
    &=-2\mathbf{M_y}^{(1)}(t)+2\mathbf{M_y}^{(1)}(t)-p\mathbf{M_y}^{(2)}(t)+2\mathbf{E}_j^\infty(\rho_{j,h})-p\mathbf{V}_j^\infty(\rho_{j,h})\\
    &\approx-p\mathbf{M_y}^{(2)}(t)+2\mathbf{E}_j^4(\rho_{j,h})-p\mathbf{V}_j^4(\rho_{j,h})
  \end{aligned}
  $}
\end{equation}
where we define:
\begin{equation}\label{vj}
    \mathbf{V}_j^\infty(\rho_{j,h}(t))=\sum_{j,h=1}^\infty j^2\rho_{j,h}(t)\frac1{2^{2h}}\\
\end{equation}
and $\mathbf{V}_j^4(\rho_{j,h})$ as the sum of the first 4 terms above.}

\textcolor{black}{Using the same reasoning, we can calculate the first moment for pure red cells. We have:
\begin{equation}\label{densred}
\resizebox{.98\hsize}{!}{$
\begin{aligned}
    \frac{d\rho_{0,i}(t)}{dt}=&-2\rho_{0,i}(t)+2\sum_{j=[\frac i2]}^{\infty}p^i\rho_{j,0}(t)\binom{2j}{i}\frac1{2^{2j}}+2\sum_{h=[\frac i2]}^{\infty}\rho_{0,h}(t)\binom{2h}{i}\frac1{2^{2h}}\\
    &+2\sum_{\substack{j+h=1\\j,h>0}}^{\infty}\rho_{j,h}(t)\binom{2h}{i}\frac1{2^{2(j+h)}}+2\sum_{\substack{j+h=[\frac{i}{2}]\\j,h>0}}^{\infty}\sum_{v=1}^i \binom{i}{v}p^v\rho_{j,h}(t)\binom{2j}{v}\binom{2h}{i-v}\frac1{2^{2(j+h)}}.
\end{aligned}
$}
\end{equation}
We can merge the 4th and 5th term together as:
\begin{equation*}
\resizebox{.98\hsize}{!}{$
\begin{aligned}
    &2\sum_{\substack{j+h=1\\j,h>0}}^{\infty}\rho_{j,h}(t)\binom{2h}{i}\frac1{2^{2(j+h)}}+2\sum_{\substack{j+h=[\frac{i}{2}]\\j,h>0}}^{\infty}\sum_{v=1}^i \binom{i}{v}p^v\rho_{j,h}(t)\binom{2j}{v}\binom{2h}{i-v}\frac1{2^{2(j+h)}}\\
    &=2\sum_{\substack{j+h=[\frac{i}{2}]\\j,h>0}}^{\infty}\sum_{v=0}^i \binom{i}{v}p^v\rho_{j,h}(t)\binom{2j}{v}\binom{2h}{i-v}\frac1{2^{2(j+h)}}
\end{aligned}    
$}
\end{equation*}
and by substituting (\ref{densred}) in (\ref{moment}) we get for the approximation with $p_u$:
\begin{equation}
\resizebox{.98\hsize}{!}{$
\begin{aligned}
    \frac{d\mathbf{M_r}^{(1)}(t)}{dt}&=\sum_{i=0}^{\infty}i\,d\rho_{0,k}(t)\\
    &=-2\mathbf{M_r}^{(1)}(t)+2\sum_{i=0}^\infty i \sum_{j=[\frac i2]}^{\infty}p^i\rho_{j,0}(t)\binom{2j}{i}\frac1{2^{2j}}+2\mathbf{M_r}^{(1)}(t)\\
    &+2\sum_{i=0}^\infty i\sum_{\substack{j+h=[\frac{i}{2}]\\j,h>0}}^{\infty}\sum_{v=0}^i \binom{i}{v}p^v\rho_{j,h}(t)\binom{2j}{v}\binom{2h}{i-v}\frac1{2^{2(j+h)}}\\
    &\approx +2p\sum_{i=0}^\infty i \sum_{j=[\frac i2]}^{\infty}\rho_{j,0}(t)\binom{2j}{i}\frac1{2^{2j}}+2p \sum_{j,h=0}^\infty \rho_{j,h}(t) \frac1{2^{2(j+h)}}\sum_{i=0}^{2(j+h)}i\sum_{v=0}^i \binom{i}{v}\binom{2j}{v}\binom{2h}{i-v}\\
    &=2p\mathbf{M_y}^{(1)}(t)+2p\mathbf{M_m}^{(1)}(t)
  \end{aligned}
  $}
\end{equation}
where we use the Vandermonde convolution formula:
\[
\sum_{v=0}^i \binom{i}{v}\binom{2j}{v}\binom{2h}{i-v}=\binom{2(j+h)}{i}.
\]
The same reasoning is used to fill Table \ref{tabmomapprox1}.}

\textcolor{black}{
Similarly to Eq. (\ref{densbluespeciessel}), we can obtain the frequency of pure yellow cells under identical positive selection, i.e. $s=s_y=s_r>1$. This is the solution of the equation:
\begin{equation}
    \frac{d\rho_{i,0}(t)}{dt}=s\bigg[\frac{d\rho_{i,0}(t)}{dt}\bigg|_{s=1}\bigg]+(s-1)\rho_{0,0}(t)\rho_{i,0}(t).
\end{equation}
By applying the formula for the moment calculation in (\ref{moment}), we get the equation for the first moment of pure yellow cells under identical positive selection in the approximation $p_u$:
\begin{equation}
\begin{aligned}
    \frac{d\mathbf{M_y}^{(1)}(t)}{dt}&=s\bigg[\frac{d\mathbf{M_y}^{(1)}(t)}{dt}\bigg]+(s-1)\mathbf{M_y}^{(1)}(t)\rho_{0,0}\\
    &=-2sp\mathbf{M}_y^{(1)}(t)+2s(1-p)\mathbf{E}_j^4(\rho_{j,h}(t))-(s-1)\mathbf{M}_y^{(1)}(t)\rho_{0,0}(t)\\
    &=-\mathbf{M}_y^{(1)}(t)\big(2sp-(s-1)\rho_{0,0}(t)\big)+2s(1-p)\mathbf{E}_j^4(\rho_{j,h}(t))
  \end{aligned}
\end{equation}
Similarly, we can get the equations for both approximations $p_l$ and $p_u$ for all the subpopulations. These are displayed in Table \ref{tabmomapprox1}.}



\begin{table}[H]
    \centering
    \resizebox{\columnwidth}{!}{%
    \begin{tabular}{ |c|c|c|c|  }
    \hline
    \multicolumn{4}{|c|}{\textbf{Weighted moment dynamics equations for $s_y=s_r=s=1$ and $p_y=p>0, p_r=0$}}\\
    \hline
    \multicolumn{4}{|c|}{First moment dynamics} \\
    \hline
    Subpopulation & & Equation with $p_l$ & Equation with $p_u$ \\
    \hline
    Total &$\frac{d\mathbf{M}^{(1)}(t)}{dt}=$  &$0$ &$0$\\
    Pure yellow &$\frac{d\mathbf{M_y}^{(1)}(t)}{dt}=$ &$-p\mathbf{M_y}^{(2)}(t)+2\mathbf{E}_j^4(\rho_{j,h})-p\mathbf{V}_j^4(\rho_{j,h})$ &$-2p\mathbf{M_y}^{(1)}(t)+2(1-p)\mathbf{E}_j^4(\rho_{j,h}(t))$\\
    Pure red &$\frac{d\mathbf{M_r}^{(1)}(t)}{dt}=$ &$2\mathbf{E}_h^4(\rho_{j,h}(t))$ &$2p\big(\mathbf{M_y}^{(1)}(t)+\mathbf{M_m}^{(1)}(t)   \big)$\\
    Mix &$\frac{d\mathbf{M_m}^{(1)}(t)}{dt}=$ &by subtraction &by subtraction\\
    \hline
\end{tabular}%
}
\caption[Equations for weighted first moment dynamics in the neutral case and one-way switching, ecDNA geno-/pheno-types]{Summary of equations for weighted first moment dynamics in the neutral case for $p_r=0$, $p_y=p$. The approximations $p_l$ and $p_u$ refers to different ways of approximating the powers of $p$ and $(1-p)$ as in (\ref{approximationfactors}).}
    \label{tabmomapprox1}
\end{table}

\begin{table}[H]
    \centering
    \resizebox{\columnwidth}{!}{%
    \begin{tabular}{ |c|c|c|c|  }
    
    \hline
    \multicolumn{4}{|c|}{\textbf{Weighted moment dynamics equations for $s_y=s_r=s>1$ and $p_y=p>0, p_r=0$}}\\
    \hline
    \multicolumn{4}{|c|}{First moment dynamics} \\
    \hline
    Subpopulation & & Equation with $p_l$ & Equation with $p_u$ \\
    \hline
    Total &$\frac{d\mathbf{M}^{(1)}(t)}{dt}=$  &$0$ &$0$\\
    Pure yellow &$\frac{d\mathbf{M_y}^{(1)}(t)}{dt}=$ &$(s-1)\rho_{0,0}\mathbf{M}_y^{(1)}(t)-sp\mathbf{M_y}^{(2)}(t)$ &$-\mathbf{M}_y^{(1)}(t)\big(2sp-(s-1)\rho_{0,0}(t)\big)$\\
    & &$+2s\mathbf{E}_j^4(\rho_{j,h})-sp\mathbf{V}_j^4(\rho_{j,h})$ &$+2s(1-p)\mathbf{E}_j^4(\rho_{j,h}(t))$\\
    Pure red &$\frac{d\mathbf{M_r}^{(1)}(t)}{dt}=$ &$(s-1)\rho_{0,0}\mathbf{M}_r^{(1)}+2s\mathbf{E}_h^4(\rho_{j,h}(t))$ &$(s-1)\rho_{0,0}\mathbf{M}_r^{(1)}+2sp\big(\mathbf{M_y}^{(1)}(t)+\mathbf{M_m}^{(1)}(t)   \big)$\\
    Mix &$\frac{d\mathbf{M_m}^{(1)}(t)}{dt}=$ &by subtraction &by subtraction\\
    \hline
\end{tabular}%
}
\caption[Equations for weighted first moment dynamics in the identical positive selection case and one-way switching, ecDNA geno-/pheno-types]{Summary of equations for weighted first moment dynamics in the identical positive selection case for $p_r=0$, $p_y=p$. The approximations $p_l$ and $p_u$ refers to different ways of approximating the powers of $p$ and $(1-p)$ as in (\ref{approximationfactors}).}
    \label{tabmom1}
\end{table}

\begin{figure}[H]
	\centering
	\includegraphics[width=\textwidth]{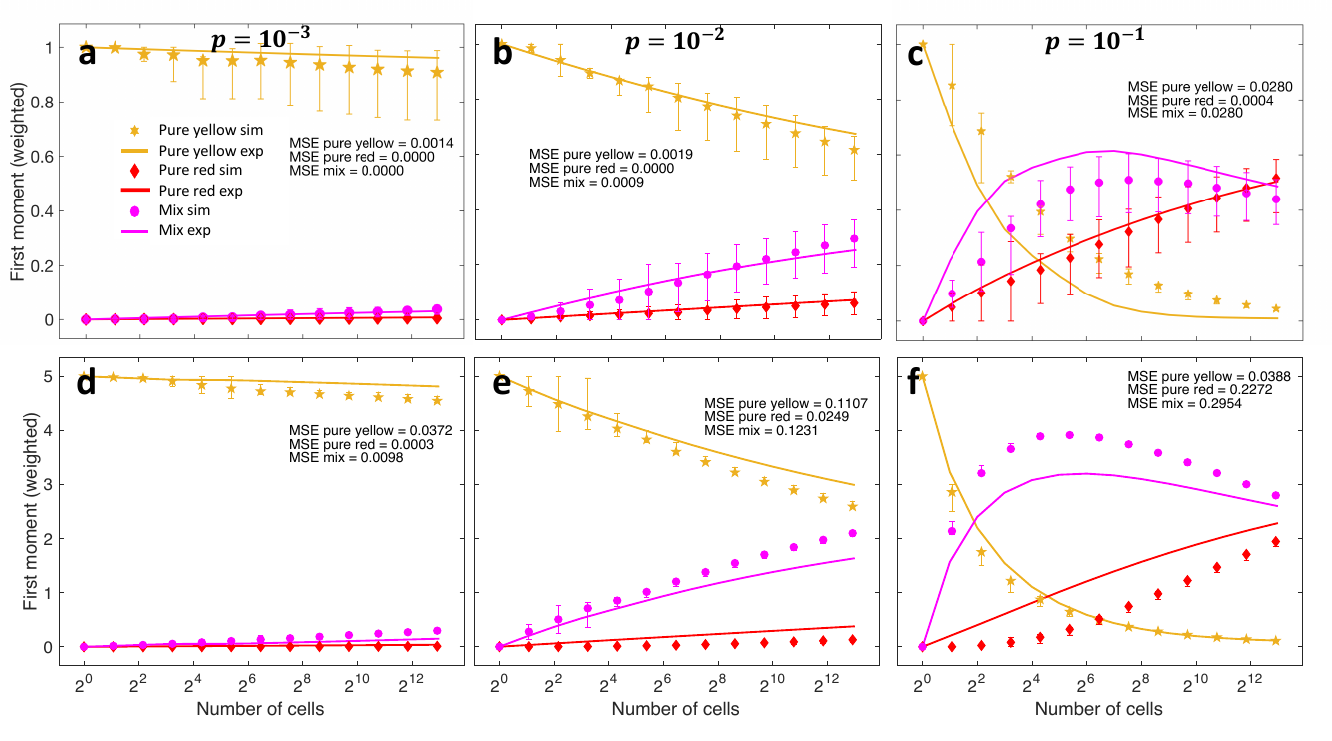}
	\caption[Weighted first moment dynamics in the neutral selection case for ecDNA geno-/pheno-types, one-way switching]{\textbf{Weighted first moment dynamics in the neutral case for one-way switching ($p_y=p$, $p_r=0$, $s_y=s_r=s=1$)}. Simulations data in comparison with upper-bound analytical approximations starting from one single pure yellow cells with 1 ecDNA copy (\textbf{a-c}), and starting from one single pure yellow cells with 5 ecDNA copies (\textbf{d-f}). Three different values for $p_y=p$ are considered; plain lines represent analytical upper-bound approximations, whilst scatter plots are data from simulations. Mean squared errors (MSE) are displayed as a measure for the distance between simulations and analytics, and 95\% confidence intervals for the mean copy numbers are shown. Results are averaged over 1000 realisations.}
	\label{F:figuramom1}
\end{figure}

However, the lower-bound equations in Table \ref{tabmomapprox1} for neutral selection (and also the ones in Supplementary Table \ref{tabmom1} for identical positive selection) do not properly account for the contribution of pure red cells in the first moment dynamics. When $p^i$ is approximated by zero for small $p$ values, it effectively nullifies the contribution due to switching to the pure red ecDNA subpopulation in the system. This limitation is not present in the upper-bound approximation, where $p^i$ approximates $p$ when $p$ is sufficiently small, ensuring that the velocity function in the ordinary differential equation for $\mathbf{M_r}^{(1)}(t)$ does account for switching. Consequently, we focus our analysis on the upper-bound approximation throughout the following sections. Our analytical approximations match well with stochastic simulations (Figure \ref{F:figuramom1} and Figure \ref{F:figuramom2}).

\begin{figure}[H]
	\centering
	\includegraphics[width=\textwidth]{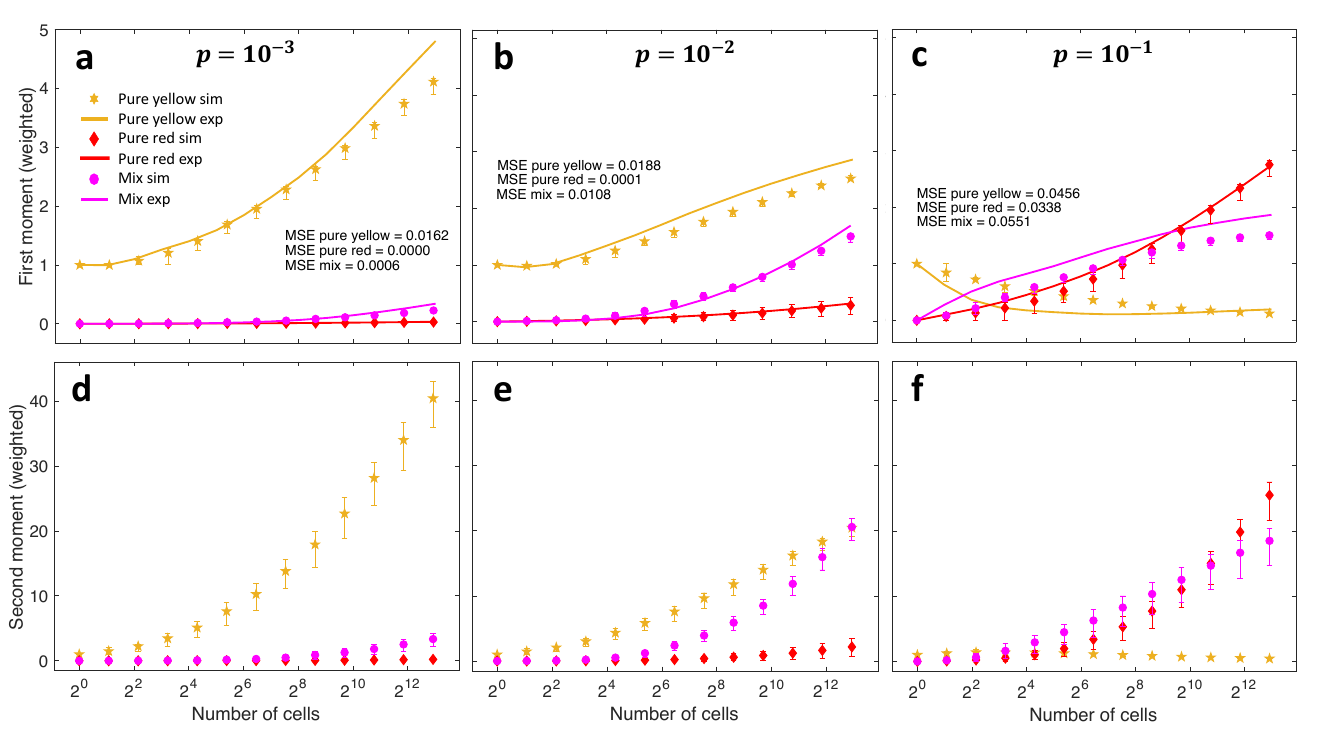}
	\caption[Weighted first and second moment dynamics in the positive selection case for ecDNA geno-/pheno-types, one-way switching]{\textbf{Weighted first and second moment dynamics in the positive selection case for one-way switching ($p_y=p>0$, $p_r=0$, $s_y=s_r=s=2$}. Simulations data in comparison with upper-bound analytical approximations for weighted first moment dynamics (\textbf{a-c}) and weighted second moment dynamics (\textbf{d-f}) of ecDNA types. Three different values for $p_y=p$ are considered; plain lines represent analytical upper-bound approximations, whilst scatter plots are data from simulations. Mean squared errors (MSE) are displayed as a measure for the distance between simulations and analytics, and 95\% confidence intervals for the mean copy numbers and variances are shown. The initial condition is a single cell with 1 yellow ecDNA copy and results are averaged over 1000 realisations.}
	\label{F:figuramomsel}
\end{figure}

Starting from a single cell and one-way switching from yellow to red ($p_y$), we examine the impact of switching on the weighted mean and variance of the copy number. When the switching rate increases, we observe a decrease of the contribution of pure yellow cells and an increase of the contribution of pure red cells to the ecDNA-positive population, no matter if under neutral selection (Figure \ref{F:figuramom1}) or identical positive selection (Figure \ref{F:figuramomsel}a-c). These considerations do not change if different initial conditions are considered (Figure \ref{F:figuramom1}d-f). With increasing switching rates, mix cells can be maintained in the population even without fitness advantages compared to pure cells. However, under neutral selection, ecDNA-positive cells will slowly decrease over time due to random segregation and the accumulation of ecDNA-free cells (Eq. \ref{rho00} and Figure \ref{F:invariant}), thus the impact of switching is stronger in this case compared to the identical positive selection scenario. When analysing small switching rates, we can appreciate indeed a difference between the two selection scenarios. Under neutral selection, the contribution of pure yellow cells decreases over time for small switching values (Figure \ref{F:figuramom1}a\&b and d\&f). On the contrary, under positive and identical selection, the ecDNA-positive cells will expand, thus, starting from a pure yellow cell, the contribution of the pure yellow subpopulation still dominates and increases when small $p$ values are considered (Figure \ref{F:figuramom2}a\&b).

The mean copy numbers align well with the weighted first moment dynamics. When selection is neutral, the absolute mean copy number sharply increases for mix cells (Figure \ref{F:meancopynumberoneway}), and we also observe a rising trend for this subpopulation in the weighted first moment (Figure \ref{F:figuramom1}), measuring the contribution of each subpopulation to the absolute mean copy number of the population.  This indicates an increase in the number of mix cells over time among the total population. On the contrary, while pure yellow ecDNA cells exhibit a constant increase in their absolute mean copy number, their decreasing trend in the weighted first moment, inversely proportional to $p$ values, suggests a decline in the number of individuals in this subpopulation over time, attributable to the one-way switching discussed above. 

Examining the variance (Figure \ref{F:figuramom2} and Figure \ref{F:figuramomsel} d-f), we observe similar dynamics to those seen in the first moment, indicating that the weighted variance in mean copy number is higher when the subpopulation is larger. This might be explained by the broad range of copy number configurations expected in a large population due to the nature of cell division.

In conclusion, we offer a few remarks on the fluctuations associated with the results presented in this section, based on the error bars shown in Figures \ref{F:figuramom1}–\ref{F:figuramomsel}. Compared to the error bars observed for ecDNA species in Figure \ref{F:figuraspecies}, here the width of the 95\% confidence intervals depends primarily on the switching parameter. We observe that lower switching values (e.g., $p=10^{-3}$) are associated with greater variability in the pure yellow compartment. Indeed, starting from a single cell carrying yellow ecDNA, the pure yellow subpopulation slowly loses cells over time due to switching and random segregation, but, since switching is rare, this variability remains largely confined to the yellow compartment. In fact, the other compartments (mix and pure red) receive cells via switching only infrequently, contributing little to their own variability. Conversely, when switching is more frequent (e.g, $p=10^{-2}-10^{-1}$), yellow ecDNA copies are more likely to switch types, increasing the probability that pure yellow cells generate daughters in other compartments. Consequently, fluctuations begin to appear in the mix and pure red compartments as well. Among these, the mix compartment still exhibits the greatest variability, as full conversion of a yellow cell into a pure red cell via switching remains a relatively rare event.

Additionally, when a higher number of initial ecDNA copies is considered, as in Figure \ref{F:figuramom1}d–f, the error bars appear less prominent due to the broader range of mean copy numbers reached by the subpopulations, even if larger bars are present at the beginning of the population expansion, when the first cells in different subpopulations appear due to the switching.

Finally, we note that fluctuations are larger under neutral selection than under positive selection (e.g., Figure \ref{F:figuramom1}a–c vs. Figure \ref{F:figuramomsel}a–c), when comparing cases with the same initial copy number. We attribute this to the fact that under positive selection, ecDNA-positive cells reach higher copy numbers more rapidly. As a result, fluctuations are proportionally smaller relative to the mean. Moreover, the reproductive advantage of ecDNA-positive cells accelerates their expansion, leading to a decline in the proportion of ecDNA-free cells. This reduces their influence on the average copy number, thereby further diminishing fluctuations.

\subsubsection*{Under identical fitness and two-way switching}

Now we investigate the case when both ecDNA types can switch into each other, i.e. $p_y, p_r\neq0$. Considering the approximations in (\ref{approximationfactors}) and the method in (\ref{metodomom}), we fill the Table \ref{tabmom2}.

\begin{table}[H]
    \centering
    \resizebox{\columnwidth}{!}{%
    \begin{tabular}{ |c|c|c|  }
    \hline
    \multicolumn{3}{|c|}{\textbf{Weighted first moment dynamics equations for $s_y=s_r=s=1$ and $p_y\neq p_r$}}\\
    \hline
    \multicolumn{3}{|c|}{First moment dynamics} \\
    \hline
    Subpopulation & & Equation (approximated) \\
    \hline
    Total &$\frac{d\mathbf{M}^{(1)}(t)}{dt}=$  &$0$\\
    Pure yellow &$\frac{d\mathbf{M_y}^{(1)}(t)}{dt}=$  &$-2p_y\mathbf{M_y}^{(1)}(t)+2p_r(1-p_y)\mathbf{M_m}^{(1)}(t)$\\
    Pure red &$\frac{d\mathbf{M_r}^{(1)}(t)}{dt}=$  &$-2p_r\mathbf{M_r}^{(1)}(t)+2p_y(1-p_r)\mathbf{M_m}^{(1)}(t)$\\
    Mix &$\frac{d\mathbf{M_m}^{(1)}(t)}{dt}=$ &\textcolor{black}{$-2p_r(1-p_y)\mathbf{M_m}^{(1)}(t)-2p_y(1-p_r)\mathbf{M_m}^{(1)}(t)+2p_y\mathbf{M_y}^{(1)}(t)+2p_r\mathbf{M_r}^{(1)}(t)$}\\
    \hline
\end{tabular}%
}
\caption[Equations for weighted first moment dynamics in the neutral case and two-way switching, ecDNA geno-/pheno-types]{Summary of equations for weighted first moment dynamics in the neutral case for $p_y\neq p_r$.}
    \label{tabmom2}
\end{table}
\textcolor{black}{Here, we ignore the contribution of pure yellow to pure red and pure red to pure yellow due to complete switching. Indeed, under small values of $p$, we can consider the entire switching of all copies almost neglectable.}

We observe an inner symmetry in the equations in Table (\ref{tabmom2}) in terms of the parameters $p_y$ and $p_r$. This, we consider an identical switching probabilities scenario for the ecDNA types, i.e. $p_y=p_r=p>0$. Focusing on the approximated equations from Table \ref{tabmom2}, we aim to solve the following system in matrix form:
\begin{equation}\label{matrixform}
    X'=AX, \qquad A=\begin{pmatrix}
-2p & 0 & 2p-2p^2 \\
0 & -2p & 2p-2p^2 \\
2p & 2p & 4p^2-4p 
\end{pmatrix}
\end{equation}
The eigenvalues of $A$ are $\lambda_1=0$, $\lambda_2=-2p$ and $\lambda_3=4p^2-6p$, with eigenvectors $v_1=(1-p, 1-p, 1)$, $v_2=(1, -1, 0)$ and $v_3=(-1, -1, 2)$. The solutions of (\ref{matrixform}) are given by
\[
\mathbf{M}=Pe^{Bt}P^{-1}C
\]
where $B$ is the diagonal similar matrix to $A$, $P$ is given by $A=PBP^{-1}$ and $C$ is the vector of initial values $C=(\mathbf{M_y}^{(1)}(0), \mathbf{M_r}^{(1)}(0), \mathbf{M_m}^{(1)}(0))$, hence:
\begin{equation*}
\begin{aligned}
  &P=\begin{pmatrix}
1-p & 1 & -1 \\
1-p & -1 & -1 \\
1 & 0 & 2 
\end{pmatrix}, \quad e^{Bt}=\begin{pmatrix}
1 & 0 & 0 \\
0 & e^{-2pt} & 0 \\
0 & 0 & e^{4p^2t-6pt} 
\end{pmatrix}, \quad P^{-1}=\begin{pmatrix}
\frac{-1}{2p-3} &\frac{-1}{2p-3}  &\frac{-1}{2p-3} \\
\frac12 & -\frac12 & -1 \\
\frac{1}{4p-6} & \frac{1}{4p-6} & \frac{p-1}{2p-3} 
\end{pmatrix}\\
\end{aligned}
\end{equation*}
and by matrix multiplication we get the solutions in Table \ref{tabsimmetry} in case of neutral selection for the whole population.

\begin{table}[H]
    \centering
    \resizebox{\columnwidth}{!}{%
    \begin{tabular}{ |c|c| }
    \hline
    \multicolumn{2}{|c|}{\textbf{Weighted first moment dynamics' for $s=s_y=s_r=1$ and $p=p_y=p_r$} (approximated)}\\
    \hline
    Subpopulation &First moment  \\
    \hline
    Total &$\mathbf{M}^{(1)}(0)$\\
    Pure yellow &$-\frac{(2(p-1)\mathbf{M_m}^{(1)}(0)+\mathbf{M_y}^{(1)}(0)+\mathbf{M_r}^{(1)}(0))e^{4p^2t-6pt}}{4p-6}$\\
    &$-\frac{\mathbf{M_r}^{(1)}(0)-\mathbf{M_y}^{(1)}(0)}{2}e^{-2pt}+\mathbf{M}^{(1)}(0)\frac{p-1}{2p-3}$\\
    Pure red &$-\frac{(2(p-1)\mathbf{M_m}^{(1)}(0)+\mathbf{M_y}^{(1)}(0)+\mathbf{M_r}^{(1)}(0))e^{4p^2t-6pt}}{4p-6}$\\
    &$+\frac{\mathbf{M_r}^{(1)}(0)-\mathbf{M_y}^{(1)}(0)}{2}e^{-2pt}+\mathbf{M}^{(1)}(0)\frac{p-1}{2p-3}$\\
    Mix & by subtraction\\
    \hline
\end{tabular}%
}
\caption[Weighted first moment dynamics in the identical selection case and two-way switching, ecDNA geno-/pheno-types]{Summary of weighted first and second moment dynamics for $s=s_r=s_y\ge1$ and $p=p_y=p_r$ (approximated).}
    \label{tabsimmetry}
\end{table}

\begin{figure}[H]
	\centering
	\includegraphics[width=\textwidth]{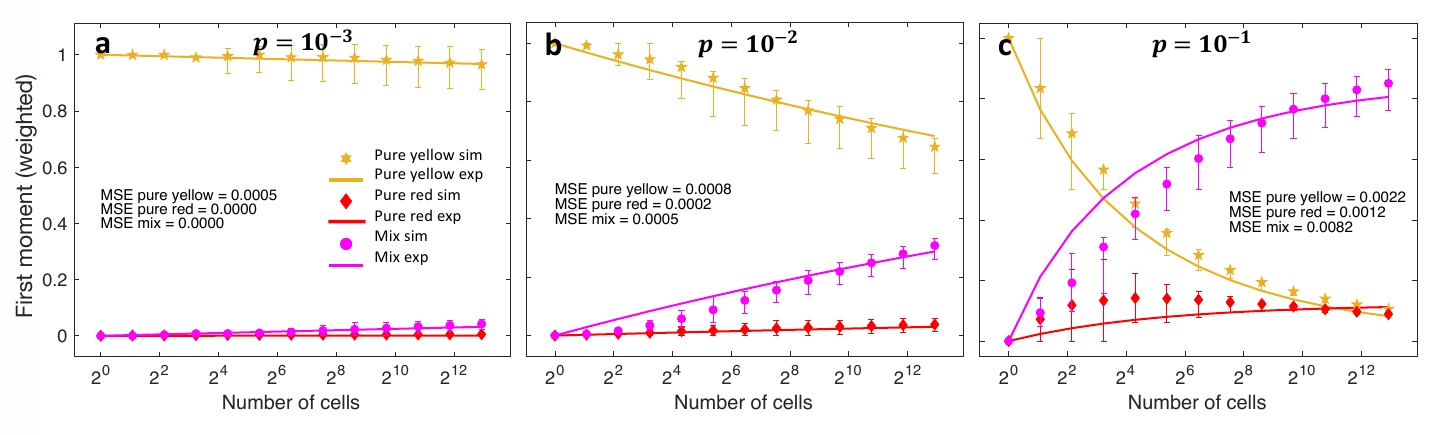}
	\caption[Weighted first moment dynamics in the neutral case for ecDNA geno-/pheno-types, two-way switching]{\textbf{Weighted first moment dynamics in the neutral selection case for two-way switching ($p_y=p_r=p$, $s_y=s_r=s=1$)}. This refers to the contribution of pure and mix cells to the mean copy number of the total population. Simulations data in comparison with upper-bound analytical approximations of first moment dynamics of ecDNA types. Three different values for $p$ are considered; plain lines represent analytical upper-bound approximations, whilst scatter plots are data from simulations. Mean squared errors (MSE) are displayed as a measure for the distance between simulations and analytics, and 95\% confidence intervals for the mean copy numbers are shown.  The initial condition is a single cell with 1 yellow ecDNA copy and results are averaged over 1000 realisations.}
	\label{F:figuraswitch}
\end{figure}

In Figure \ref{F:figuraswitch}, starting from a single cell with one yellow ecDNA copy and a two-way switching from yellow to red copies and vice-versa, we observe a consistent alignment between our analytical approximations and stochastic simulations under neutral selection. Comparing the mean copy number in Figure \ref{F:meancopynumbertwoway} with the one-switching case in Figure \ref{F:meancopynumberoneway}, we observe a more pronounced increase in mix cell trends for both absolute and relative counts when $p$ is relatively high. Furthermore, the absolute mean copy number for pure cells appears to reach a sort of steady state after several generations, suggesting that mix cells rapidly become the ones with the highest number of copies in the two-way switching scenario.

Additionally, by examining the weighted mean copy number in Figure \ref{F:figuraswitch}, we can draw additional conclusions about the proportion of subpopulations over time. Pure yellow cells decrease in number over time, while mix cells significantly increase. Summarising the results from both absolute and weighted first moments, it appears that mix cells are increasingly dominating the ecDNA-positive cells, carrying also the configurations with the highest copy numbers. Comparing these observations with the ones under one-way switching (Figure \ref{F:figuramom1}), it seems that two-way switching leads to a sharper increase of weighted mean copy number of mix. This suggests the two way switching promote the expansion and maintainance of multiple ecDNA types presenting in the same cells.

\section*{Discussion}

We present a framework to model two ecDNA types, which can be adapted to either species, genotypes or phenotypes, by easily changing the switching probabilities between the two types.  If the type switching probability is set to zero for both yellow and red ecDNA, we consider two distinct species; conversely, we consider two genotypes distinct by point mutations (with relatively low switching probabilities) or phenotypes by phenotypical switching. Similar to previous models on a single ecDNA type \cite{22}, our investigation demonstrates that random segregation of ecDNA in cell divisions significantly impacts the system's heterogeneity, discernible through the identification of multiple subpopulations of cells carrying or lacking specific ecDNA types. Two evolutionary parameters, namely the fitness coefficient and type switching probability, are central to our analytical exploration. 

The distribution of ecDNA copies within the population and the densities of subpopulations over time are influenced by those parameters. While the mean copy number of the population will increase with more cell divisions if fitness is positive, it takes some time for the ecDNA to spread in a growing population. Thus, in simulations the initial configuration of cells also largely determines the copy number distribution in observation. A lower initial ecDNA copy number results in a distribution with a sharper peak and a lower mean copy number across the population (Figure \ref{F:invariant}c\&d, curve representing 1 yellow and 1 red ecDNA copies as the initial configuration). \textcolor{black}{A higher initial copy number yields a distribution with a more bell-shaped curve, indicative of an overall higher mean copy number (Figure \ref{F:invariant}c\&d, curve representing 10 yellow and 10 red ecDNA copies as the initial configuration). In reality, while ecDNA is most likely to arise with a single copy, the expansion of ecDNA-positive cells might only start with a cell carrying multiple copies. In addition, giving sufficient long enough time, the ecDNA copy number distributions will evolve to a more bell-shaped curve as observed more often in cell lines \cite{22}.} 

As expected when both yellow and red ecDNA have identical fitness either neutral or positive,  the whole population has the same behaviour as it has a single type of ecDNA \cite{22}. Variations in type switching probability ($p$) do not impact the distribution of copies in ecDNA-positive cells or the density of ecDNA-free cells over time.  However, once one of the ecDNA types carries a different reproduction rate, the independence of these quantities from the type switching probability disappears. Moreover, if the switching is identical and two-way,  intermediate values of $p$ lead to higher mean copy numbers and a greater proportion of cells carrying ecDNA (Figure \ref{F:invariant}b\&d). Notably, the distribution in Figure \ref{F:invariant}d exhibits increased width, while the fraction of ecDNA-free cells in Figure \ref{F:invariant}b diminishes for values of $p$ approaching 0.5. Overall, this observation suggests a potential advantage in maintaining mix cells over time, facilitated by intermediate values of $p$. The presence of type switching enables pure mother cells to produce mix daughters, and the efficacy of this process seems most pronounced when $p$ assumes intermediate values. This hypothesis looks intriguing and will be further investigated in the next chapter.

We analysed the mean and variance of distinct subpopulations through the first and second moment dynamics for both ecDNA species and geno-/pheno-types. In the context of ecDNA species, selection affects the mean ecDNA copy number for cells carrying either just yellow or just red copies, resulting in a substantial increase over time. Considering then ecDNA geno-/pheno-types, we categorised our analysis on various parameter choices using Taylor approximation for small values of type switching probability. The solutions to the moment dynamics provided align well with simulations for small probability values but may exhibit significant discrepancies for larger values. Nevertheless, small switching probabilities are biologically more relevant for our model \cite{140, 105}.

If identical fitness and one-way switching is considered, for example from yellow to red, the mean and variation for the pure yellow population exhibit a progressive decrease with an increase in the switching probability. Conversely, the opposite trend is observed for the pure red population. This is observable in different ways. First by looking at the analytical approximations in Table \ref{tabmom1}, the normalised average copy number for pure red cells, i.e. their first moment, is increasing, compared to ones for pure yellow, that decreases proportionally to $\approx e^{-pt}$, and mix cells; the variance is accordingly monotonically increasing for pure red. Secondly, we present how first and second moments changes over time in simulations in Figure \ref{F:figuramom1}, which agrees to our analytical results above. These observations suggest that in the case of one-way switching from yellow to red, under conditions of identical positive fitness for cells carrying ecDNA, pure red cells are going to be the majority on the long run. The opposite behaviour is expected if the one-way switching is considered from red to yellow ecDNA.

However, if the switching is two-way between the two ecDNA types, the mix population shows an increase in the average copy number proportionate to the switching probability. This suggests that intermediate levels of switching may create advantageous conditions for the proliferation of mix cells, paving the way for further investigations into the optimal switching dynamics.

In summary, we constructed a general theoretical model of multiple ecDNA types, which match the recent experimental and clinical observations and provide a general framework to investigate the complex dynamics of extra-chromosomal DNA interactions in cancers.

\section*{Acknowledgements}

E.S. is supported by the Engineering and Physical Sciences Research Council (EPSRC) PhD fellowship that is part of the United Kingdom Research and Innovation association (UKRI). 
B.W. is supported by a Barts Charity Lectureship (grant no. MGU045) and a UKRI Future Leaders Fellowship (grant no. MR/V02342X/1).\\
W.H. is funded by NNSF General Program (grant no. 3217024). 
E.S. and W.H. are funded by Cancer Grand Challenges CGCSDF-2021\textbackslash{}100007 with support from Cancer Research UK and the National Cancer Institute.

\bibliographystyle{plain} 
\bibliography{Bibliography} 

\begin{thebibliography}{10}

\bibitem{5}
L'Abbate A, Macchia G, D'Addabbo P, Lonoce A, Tolomeo D, Trombetta D, Kok K, Bartenhagen C, Whelan CW, Palumbo O, Severgnini M, Cifola I, Dugas M, Carella M, De~Bellis G, Rocchi M, Carbone L, and Storlazzi CT.
\newblock Genomic organization and evolution of double minutes/homogeneously staining regions with {MYC} amplification in human cancer.
\newblock {\em Nucleic Acids Research}, 42:9131–9145, 2014.

\bibitem{3}
Bailey C, Shoura MJ, Mischel PS, and Swanton C.
\newblock Extrachromosomal {DNA}-relieving heredity constraints, accelerating tumour evolution.
\newblock {\em Annals of Oncology}, 31:884--893, 2020.

\bibitem{38}
Bailey C, Shoura MJ, Mischel PS, and Swanton C.
\newblock Extrachromosomal {DNA}—relieving heredity constraints, accelerating tumour evolution.
\newblock {\em Annals of Oncology}, 31(7):884--893, 2020.

\bibitem{4}
Aktipis CA, Boddy AM, Gatenby RA, Brown JS, and Maley CC.
\newblock Life history trade-offs in cancer evolution.
\newblock {\em Nature Reviews Cancer}, 13:883--892, 2013.

\bibitem{27}
Mas-Ponte D and Supek F.
\newblock {DNA} mismatch repair promotes {APOBEC3}-mediated diffuse hypermutation in human cancers.
\newblock {\em Nature Genetics}, 52:958--968, 2020.

\bibitem{17}
Nathanson DA, Gini B, Mottahedeh J, Visnyei K, Koga T, Gomez G, Eskin A, Hwang K, Wang J, Masui K, Paucar A, Yang H, Ohashi M, Zhu S, Wykosky J, Reed R, Nelson SF, Cloughesy TF, James CD, Rao PN, Kornblum HI, Heath JR, Cavenee WK, Furnari FB, and Mischel PS.
\newblock Targeted therapy resistance mediated by dynamic regulation of extrachromosomal mutant {EGFR DNA}.
\newblock {\em Science}, 343:72--76, 2014.

\bibitem{33}
Gillespie DT.
\newblock A general method for numerically simulating the stochastic time evolution of coupled chemical reactions.
\newblock {\em Journal of Computational Physics}, 22(4):403--434, 1976.

\bibitem{34}
Gillespie DT.
\newblock Exact stochastic simulation of coupled chemical reactions.
\newblock {\em The Journal of Physical Chemistry}, 81(25):2340--2361, 1977.

\bibitem{35}
Gillespie DT.
\newblock Stochastic simulation of chemical kinetics.
\newblock {\em Annual Review of Physical Chemistry}, 58:35--55, 2007.

\bibitem{105}
Scanu E, Werner B, and Huang W.
\newblock Population dynamics of multiple {ecDNA} types.
\newblock {\em in preparation}, 2024.

\bibitem{23}
Yi~E, Gujar AD, Guthrie M, Kim H, Zhao D, Johnson KC, Amin SB, Costa ML, Yu~Q, Das S, Jillette N, Clow PA, Cheng AW, and Verhaak RGW.
\newblock Live-cell imaging shows uneven segregation of extrachromosomal {DNA} elements and transcriptionally active extrachromosomal {DNA} hubs in cancer.
\newblock {\em Cancer Discovery}, 12:468--483, 2022.

\bibitem{19}
Yi~E, Chamorro~González R, Henssen AG, and Verhaak RGW.
\newblock Extrachromosomal {DNA} amplifications in cancer.
\newblock {\em Nature Reviews Genetics}, 23:760--771, 2022.

\bibitem{28}
Bergstrom EN, Luebeck J, Petljak M, Khandekar A, Barnes M, Zhang T, Steele CD, Pillay N, Landi MT, Bafna V, Mischel PS, Harris RS, and Alexandrov LB.
\newblock Mapping clustered mutations in cancer reveals {APOBEC3} mutagenesis of {ecDNA}.
\newblock {\em Nature}, 602:510--517, 2022.

\bibitem{Cumming2007}
Cumming G, Fidler F, and Vaux DL.
\newblock Error bars in experimental biology.
\newblock {\em Journal of Cell Biology}, 177(1):7--11, 2007.

\bibitem{25}
Stark GR, Debatisse M, Giulotto E, and Wahl GM.
\newblock Recent progress in understanding mechanisms of mammalian {DNA} amplification.
\newblock {\em Cell}, 57:901--908, 1989.

\bibitem{8}
Kim H, Nguyen NP, Turner K, Wu~S, Gujar AD, Luebeck J, Liu J, Deshpande V, Rajkumar U, Namburi S, Amin SB, Yi~E, Menghi F, Schulte JH, Henssen AG, Chang HY, Beck CR, Mischel PS, Bafna V, and Verhaak RGW.
\newblock Extrachromosomal {DNA} is associated with oncogene amplification and poor outcome across multiple cancers.
\newblock {\em Nature Genetics}, 52:891--897, 2020.

\bibitem{140}
Martincorena I and Peter JC.
\newblock Somatic mutation in cancer and normal cells.
\newblock {\em Science}, 349.625:1483--1489, 2015.

\bibitem{Jones2}
Johnston IG, Burgstaller JP, Havlicek V, Kolbe T, Rülicke T, Brem G, Poulton J, and Jones NS.
\newblock Stochastic modelling, bayesian inference, and new in vivo measurements elucidate the debated {mtDNA} bottleneck mechanism.
\newblock {\em Elife}, 4:e07464, 2015.

\bibitem{Jones1}
Johnston IG and Jones NS.
\newblock Closed-form stochastic solutions for non-equilibrium dynamics and inheritance of cellular components over many cell divisions.
\newblock {\em Proceedings of the Royal Society A: Mathematical, Physical and Engineering Sciences}, 471(2180):20150050, 2015.

\bibitem{31}
Rose JC, Wong ITL, Daniel B, Jones MG, Yost KE, Hung KL, Curtis EJ, Mischel PS, and Chang HY.
\newblock Disparate pathways for extrachromosomal {DNA} biogenesis and genomic {DNA} repair.
\newblock {\em bioRxiv}, 2023.

\bibitem{20}
Lange JT, Chen CY, Pichugin Y, Xie, Tang J, Hung KL, Yost KE, Shi Q, Erb ML, Rajkumar U, Wu~S, Swanton C, Liu Z, Huang W, Chang HY, Bafna V, Henssen AG, Werner B, and Mischel PS.
\newblock Principles of {ecDNA} random inheritance drive rapid genome change and therapy resistance in human cancers.
\newblock {\em bioRxiv}, 2021.

\bibitem{22}
Lange JT, Rose JC, Chen CY, Pichugin Y, Xie L, Tang J, Hung KL, Yost KE, Shi Q, Erb ML, Rajkumar U, Wu~S, Taschner-Mandl S, Bernkopf M, Swanton C, Liu Z, Huang W, Chang HY, Bafna V, Henssen AG, Werner B, and Mischel PS.
\newblock The evolutionary dynamics of extrachromosomal {DNA} in human cancers.
\newblock {\em Nature Genetics}, 54:1527--1533, 2022.

\bibitem{6}
Sanborn JZ, Salama SR, Grifford M, Brennan CW, Mikkelsen T, Jhanwar S, Katzman S, Chin L, and Haussler D.
\newblock Double minute chromosomes in glioblastoma multiforme are revealed by precise reconstruction of oncogenic amplicons.
\newblock {\em Cancer Research}, 73:6036--6045, 2013.

\bibitem{24}
Song K, Minami JK, Huang A, Dehkordi SR, Lomeli SH, Luebeck J, Goodman MH, Moriceau G, Krijgsman O, Dharanipragada P, Ridgley T, Crosson WP, Salazar J, Pazol E, Karin G, Jayaraman R, Balanis NG, Alhani S, Sheu K, Ten~Hoeve J, Palermo A, Motika SE, Senaratne TN, Paraiso KH, Hergenrother PJ, Rao PN, Multani AS, Peeper DS, Bafna V, Lo~RS, and Graeber TG.
\newblock Plasticity of extrachromosomal and intrachromosomal {BRAF} amplifications in overcoming targeted therapy dosage challenges.
\newblock {\em Cancer Discovery}, 12:1046--1069, 2022.

\bibitem{26}
Johnson KC, Anderson KJ, Courtois ET, Gujar AD, Barthel FP, Varn FS, Luo D, Seignon M, Yi~E, Kim H, Estecio MRH, Zhao D, Tang M, Navin NE, Maurya R, Ngan CY, Verburg N, De~Witt~Hamer PC, Bulsara K, Samuels ML, Das S, Robson P, and Verhaak RGW.
\newblock Single-cell multimodal glioma analyses identify epigenetic regulators of cellular plasticity and environmental stress response.
\newblock {\em Nature Genetics}, 53:1456--1468, 2021.

\bibitem{36}
Hung KL, Luebeck J, Dehkordi SR, Colón CI, Li~R, Wong IT, Coruh C, Dharanipragada P, Lomeli SH, Weiser NE, Moriceau G, Zhang X, Bailey C, Houlahan KE, Yang W, González RC, Swanton C, Curtis C, Jamal-Hanjani M, Henssen AG, Law JA, Greenleaf WJ, Lo~RS, Mischel PS, Bafna V, and Chang HY.
\newblock Targeted profiling of human extrachromosomal {DNA} by {CRISPR}-{CATCH}.
\newblock {\em Nature Genetics}, 54:1746--1754, 2022.

\bibitem{16}
Hung KL, Yost KE, Xie L, Shi Q, Helmsauer K, Luebeck J, Schöpflin R, Lange JT, Chamorro~González R, Weiser NE, Chen C, Valieva ME, Wong IT, Wu~S, Dehkordi SR, Duffy CV, Kraft K, Tang J, Belk JA, Rose JC, Corces MR, Granja JM, Li~R, Rajkumar U, Friedlein J, Bagchi A, Satpathy AT, Tjian R, Mundlos S, Bafna V, Henssen AG, Mischel PS, Liu Z, and Chang HY.
\newblock {ecDNA} hubs drive cooperative intermolecular oncogene expression.
\newblock {\em Nature}, 600:731--736, 2021.

\bibitem{29}
Hung KL, Jones MG, Wong IT, Lange JT, Luebeck J, Scanu E, He~BJ, Brückner L, Li~R, González RC, Schmargon R, Dörr JR, Belk JA, Bafna V, Werner B, Huang W, Henssen AG, Mischel PS, and Chang HY.
\newblock Coordinated inheritance of extrachromosomal {DNA} species in human cancer cells.
\newblock {\em bioRxiv}, 2023.

\bibitem{39}
Hung KL, Mischel PS, and Chang HY.
\newblock Gene regulation on extrachromosomal {DNA}.
\newblock {\em Nature Structural and Molecular Biology}, 29:736–744, 2022.

\bibitem{14}
Turner KM, Deshpande V, Beyter D, Koga T, Rusert J, Lee C, Li~B, Arden K, Ren B, Nathanson DA, Kornblum HI, Taylor MD, Kaushal S, Cavenee WK, Wechsler-Reya R, Furnari FB, Vandenberg SR, Rao PN, Wahl GM, Bafna V, and Mischel PS.
\newblock Extrachromosomal oncogene amplification drives tumour evolution and genetic heterogeneity.
\newblock {\em Nature}, 543:122--125, 2017.

\bibitem{21}
Turner KM, Deshpande V, Beyter D, Koga T, Rusert J, Lee C, Li~B, Arden K, Ren B, Nathanson DA, Kornblum HI, Taylor MD, Kaushal S, Cavenee WK, Wechsler-Reya R, Furnari FB, Vandenberg SR, Rao PN, Wahl GM, Bafna V, and Mischel PS.
\newblock Extrachromosomal oncogene amplification drives tumour evolution and genetic heterogeneity.
\newblock {\em Nature}, 543:122--125, 2017.

\bibitem{37}
Pecorino LT, Verhaak RGW, Henssen A, and Mischel PS.
\newblock Extrachromosomal {DNA} ({ecDNA}): an origin of tumor heterogeneity, genomic remodeling, and drug resistance.
\newblock {\em Biochemical Society Transactions}, 50(6):1911–1920, 2022.

\bibitem{2}
McGranahan N and Swanton C.
\newblock Clonal heterogeneity and tumor evolution: Past, present, and the future.
\newblock {\em Cell}, 168(4):613--628, 2017.

\bibitem{13}
Verhaak RG, Bafna V, and Mischel PS.
\newblock Extrachromosomal oncogene amplification in tumour pathogenesis and evolution.
\newblock {\em Nature Reviews Cancer}, 19:283--288, 2019.

\bibitem{32}
Nakaoka S and Aihara K.
\newblock Stochastic simulation of structured skin cell population dynamics.
\newblock {\em Journal of Mathematical Biology}, 66:807--835, 2013.

\bibitem{15}
Wu~S, Turner KM, Nguyen N, Raviram R, Erb M, Santini J, Luebeck J, Rajkumar U, Diao Y, Li~B, Zhang W, Jameson N, Corces MR, Granja JM, Chen X, Coruh C, Abnousi A, Houston J, Ye~Z, Hu~R, Yu~M, Kim H, Law JA, Verhaak RGW, Hu~M, Furnari FB, Chang HY, Ren B, Bafna V, and Mischel PS.
\newblock Circular {ecDNA} promotes accessible chromatin and high oncogene expression.
\newblock {\em Nature}, 575:699--703, 2019.

\bibitem{1}
Wu~S, Bafna V, and Mischel PS.
\newblock Extrachromosomal {DNA} ({ecDNA}) in cancer pathogenesis.
\newblock {\em Current Opinion in Genetics and Development}, 66:78--82, 2021.

\bibitem{18}
Liao Z, Jiang W, Ye~L, Li~T, Yu~X, and Liu L.
\newblock Classification of extrachromosomal circular {DNA} with a focus on the role of extrachromosomal {DNA} ({ecDNA}) in tumor heterogeneity and progression.
\newblock {\em Biochimica et Biophysica Acta Reviews on Cancer}, 1874(1), 2020.

\end{thebibliography}

\section*{Supplementary Figures}

\setcounter{equation}{0}
\setcounter{figure}{0}
\setcounter{table}{0}
\makeatletter
\renewcommand{\theequation}{S\arabic{equation}}
\renewcommand{\thefigure}{S\arabic{figure}}
\renewcommand{\thetable}{S\arabic{table}}

\begin{figure}[H]
	\centering
	\includegraphics[width=0.8\textwidth]{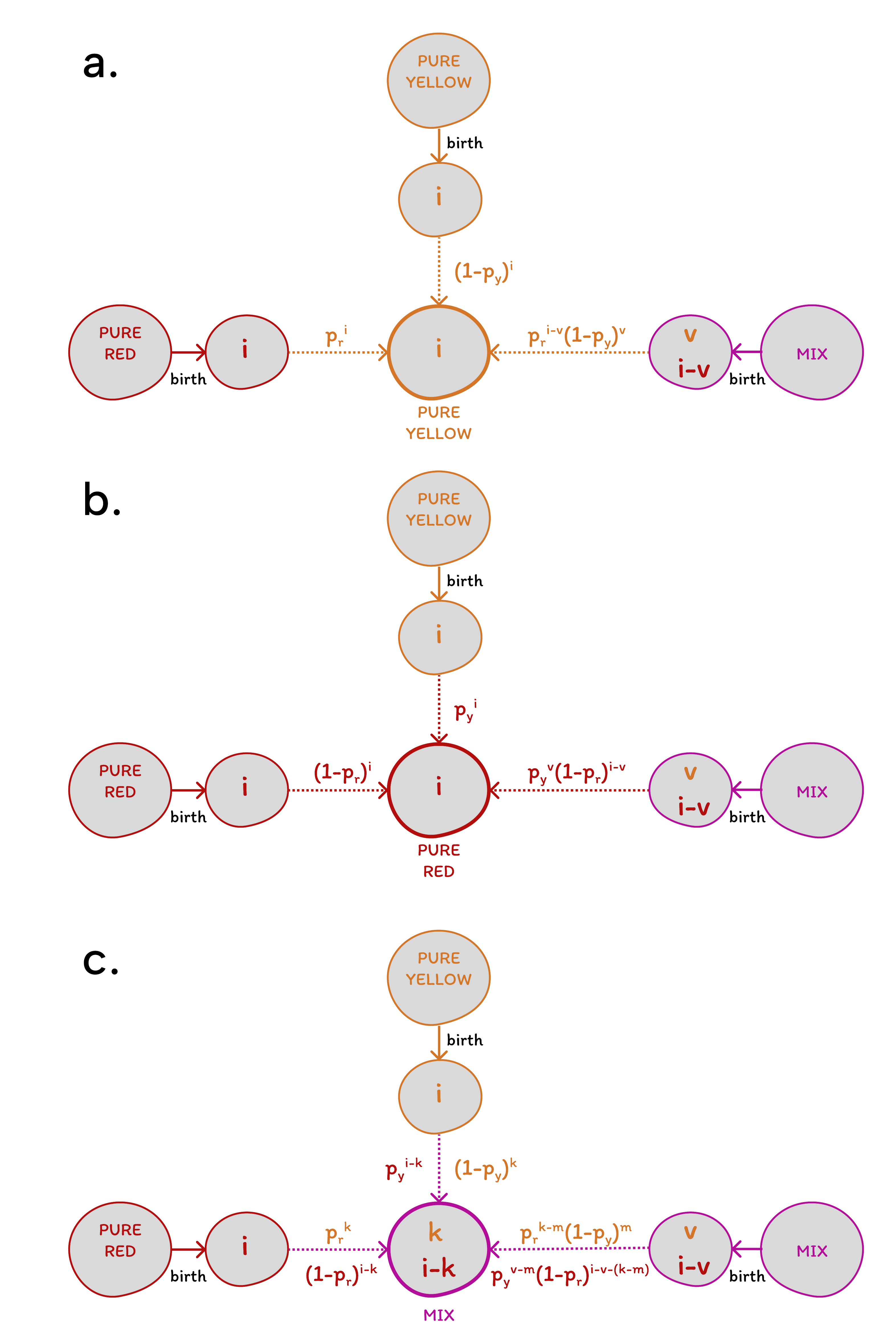}
	\caption[Illustration of equations for the dynamical system]{\textbf{Illustration of equations for the dynamical system.} We illustrate the interactions among subpopulations as in the system of equations (\ref{sistemaswitch}). \textbf{a.} Equation for pure yellow cells, i.e. $C_{i,0}(t)$. \textbf{b.} Equation for pure red cells, i.e. $C_{0,i}(t)$. \textbf{c.} Equation for mix cells, i.e. $C_{k,i-k}(t)$(t) transitions (positive terms) due to the switching, represented by the possibility of recruiting pure yellow, pure red and mix cells (from another combination $\neq (k,i-k)$), and the "out" transitions (negative terms) due to the switching, represented by the loss of mix $(k,i-k)$ cells in favour of pure yellow, pure red and mix ones (with another combination $\neq (k,i-k)$).}
	\label{equation}
\end{figure}

\begin{figure}[H]
	\centering
	\includegraphics[width=\textwidth]{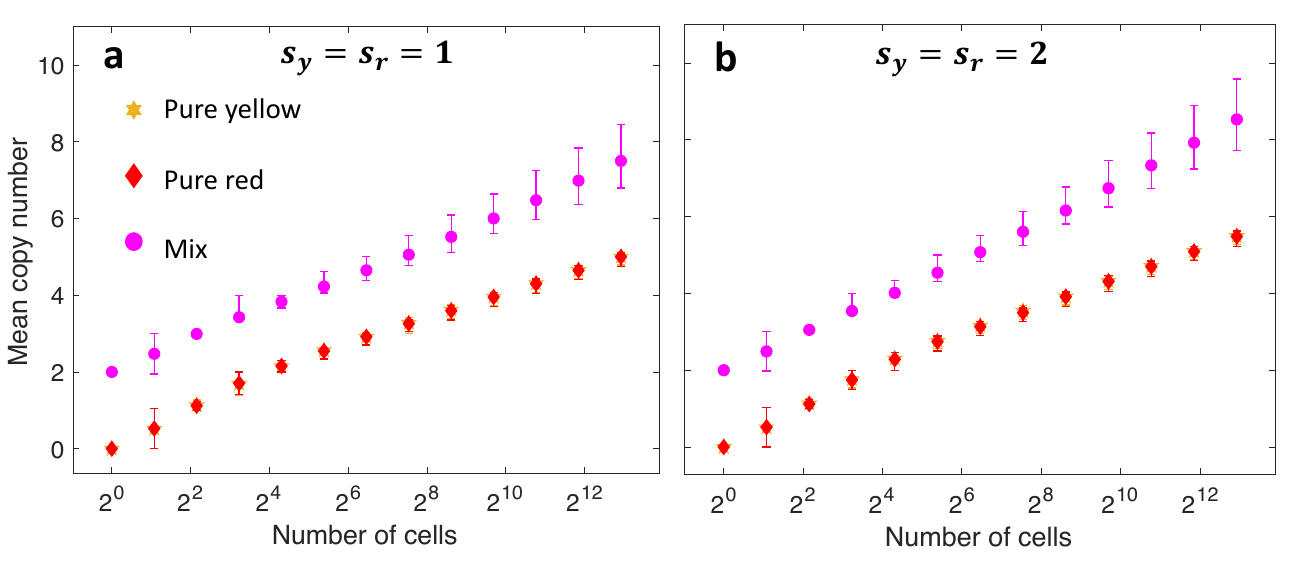}
	\caption[Mean copy number for ecDNA species]{\textbf{Mean copy number for ecDNA species ($p_y=p_r=p=0$)}. We show simulations data for mean copy number for ecDNA species under different selection scenarios and initial conditions. \textbf{a.} Neutral selection \textbf{b.} Identical positive selection. In both panels, 95\% confidence intervals for the mean copy numbers are shown. The initial condition is a single cell with 1 yellow and 1 red ecDNA copy, and this is why the trends for pure yellow and pure red ecDNA are coinciding. Results are averaged over 1000 realisations.}
	\label{F:meancopynumberspecies}
\end{figure}

\begin{figure}[H]
	\centering
	\includegraphics[width=0.9\textwidth]{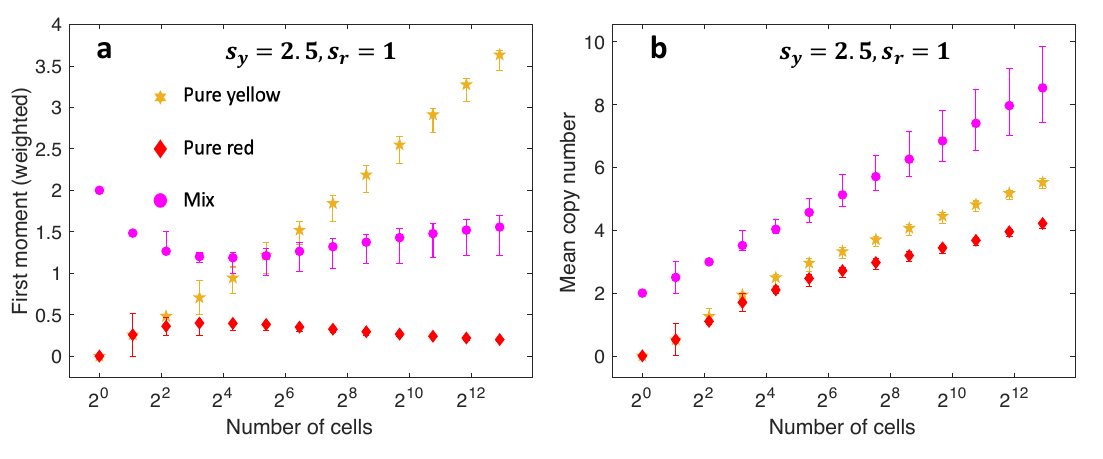}
	\caption[First moment dynamics for ecDNA species under non-identical fitness scenario]{\textbf{First moment dynamics for ecDNA species under non-identical fitness scenario ($p_y=p_r=p=0$, $s_y=2.5, s_r=1$).} \textbf{a. Weighted first moment.} This refers to the contribution of pure and mix cells to the mean copy number of the total population. \textbf{b. Mean copy number.} This refers to the absolute mean copy number of the subpopulations. In both panels, 95\% confidence intervals for the mean copy numbers are shown. The initial condition is a single cell with 1 yellow and 1 red ecDNA copy and results are averaged over 1000 realisations.}
	\label{F:figuraspeciesnonid}
\end{figure}

\begin{figure}[H]
	\centering
	\includegraphics[width=\textwidth]{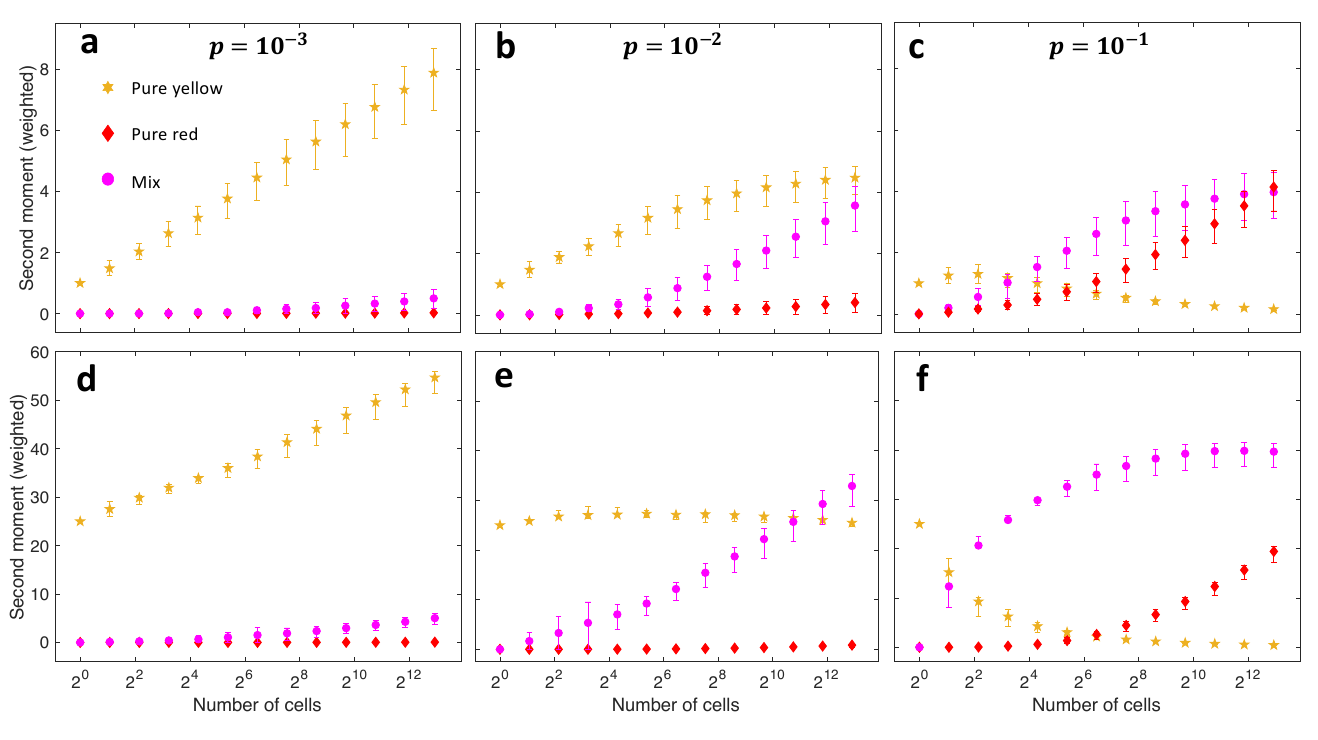}
	\caption[Weighted second moment dynamics in the neutral selection case for ecDNA geno-/pheno-types, one-way switching]{\textbf{Weighted second moment dynamics in the neutral case for one-way switching ($p_y=p$, $p_r=0$, $s_y=s_r=s=1$)}. We show weighted second moment dynamics for each subpopulation starting from one single pure yellow cells with 1 ecDNA copy (\textbf{a-c}), and starting from one single pure yellow cells with 5 ecDNA copies (\textbf{d-f}). Three different values for $p_y=p$ are considered. 95\% confidence intervals for the mean variances are shown. Results are averaged over 1000 realisations.}
	\label{F:figuramom2}
\end{figure}

\begin{figure}[H]
	\centering
	\includegraphics[width=\textwidth]{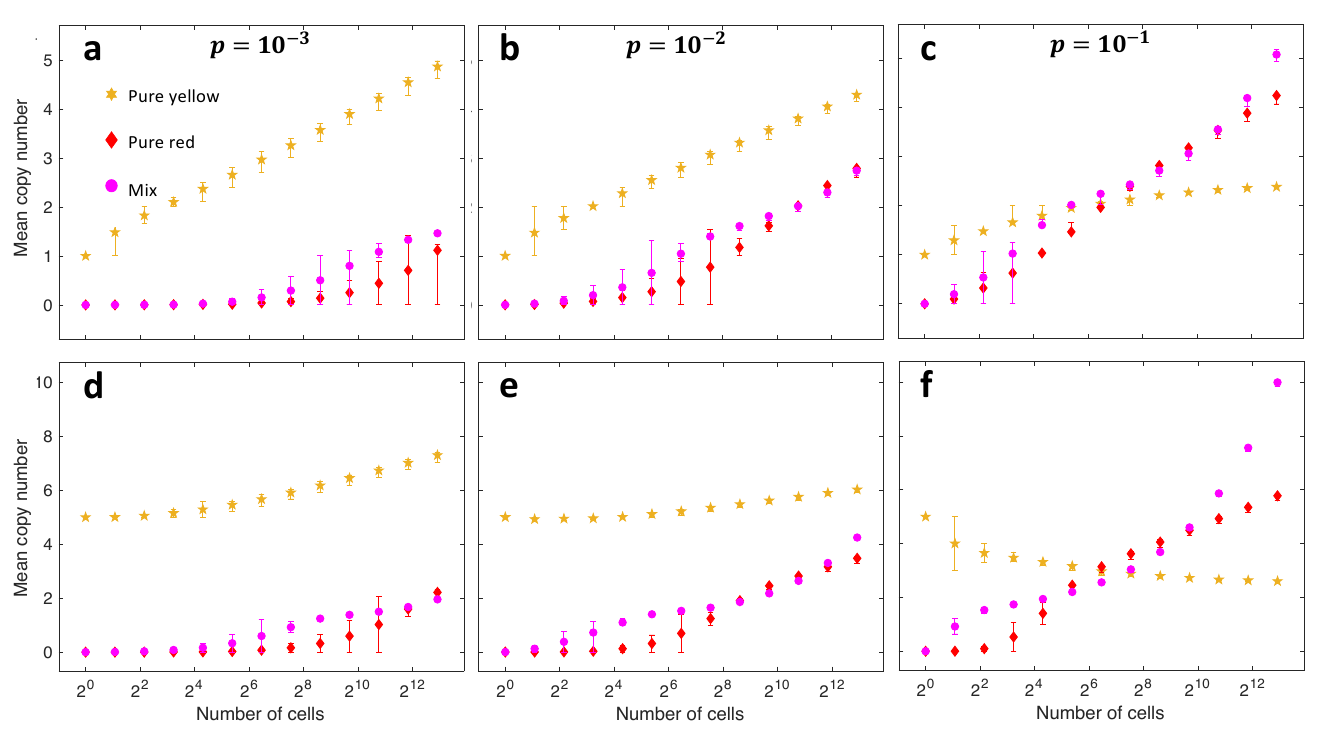}
	\caption[Mean copy number in the neutral selection case for ecDNA geno-/pheno-types, one-way switching]{\textbf{Mean copy number under neutral selection for one-way switching ($p_y=p$, $p_r=0$, $s_y=s_r=s=1$)}. We show mean copy number for each subpopulation starting from one single pure yellow cells with 1 ecDNA copy (\textbf{a-c}), and starting from one single pure yellow cells with 5 ecDNA copies (\textbf{d-f}). Three different values for $p_y=p$ are considered. 95\% confidence intervals for the mean copy numbers are shown. Results are averaged over 1000 realisations.}
	\label{F:meancopynumberoneway}
\end{figure}

\begin{figure}[H]
	\centering
	\includegraphics[width=\textwidth]{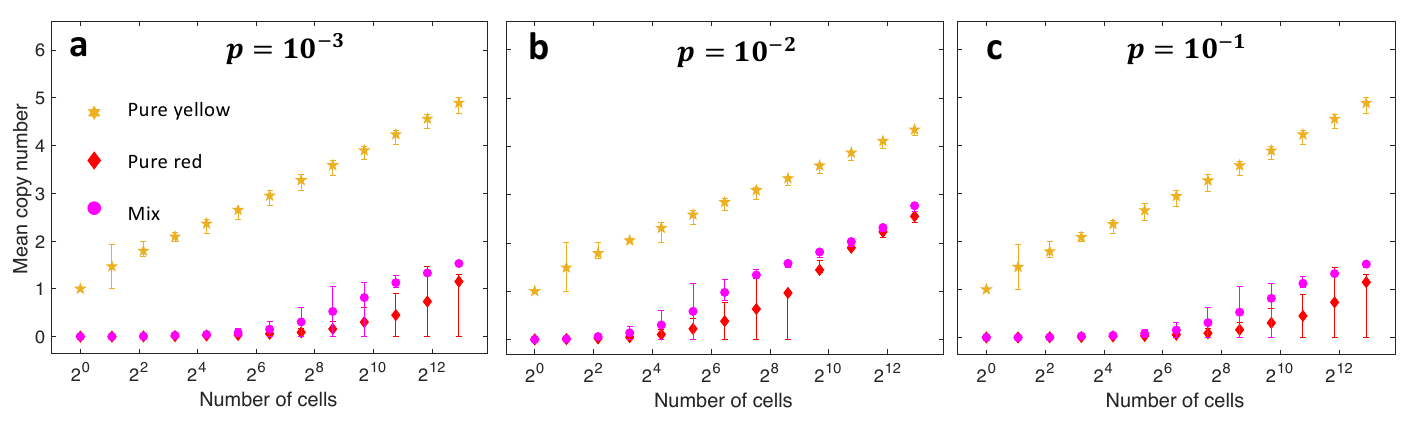}
	\caption[Mean copy number in the neutral case for ecDNA geno-/pheno-types, two-way switching]{\textbf{Mean copy number under neutral selection for two-way switching ($p_y=p_r=p$, $s_y=s_r=s=1$)}. We show simulations data for mean copy number for ecDNA types under neutral selection and identical two-way switching. Three different values for $p_y=p$ are considered. 95\% confidence intervals for the mean copy numbers are shown. The initial condition is a single cell with 1 yellow ecDNA copy and results are averaged over 1000 realisations.}
	\label{F:meancopynumbertwoway}
\end{figure}

\end{document}